\documentclass[prb,twocolumn,superscriptaddress,floatfix,letterpaper]{revtex4-1}
\pdfoutput=1

\usepackage[all]{xy}

\usepackage{euscript}
\newcommand\Rep{\EuScript{R}\mathrm{ep}}
\newcommand\sRep{\mathrm{s}\EuScript{R}\mathrm{ep}}

\newcommand\sC{\EuScript{C}}
\newcommand\sD{\EuScript{D}}
\newcommand\sE{\EuScript{E}}

\DeclareMathOperator{\Hom}{Hom}
\DeclareMathOperator{\id}{id}

\DeclareMathOperator{\Aut}{Aut}

\newcommand{\Ve}{\mathrm{Vect}}

\begin{document}

\title{A classification of 3+1D bosonic topological orders (I):\\
the case when point-like excitations are all bosons
}

\author{Tian Lan} 
\affiliation{Perimeter Institute for Theoretical Physics, Waterloo, Ontario N2L 2Y5, Canada} 
\affiliation{Institute for Quantum Computing,
  University of Waterloo, Waterloo, Ontario N2L 3G1, Canada}

\author{Liang Kong} 
\affiliation{Yau Mathematical Science Center, Tsinghua University, Beijing, 100084, China}

\author{Xiao-Gang Wen}
\affiliation{Department of Physics, Massachusetts Institute of
Technology, Cambridge, Massachusetts 02139, USA}

\begin{abstract} 
Topological orders are new phases of matter beyond Landau symmetry breaking.
They correspond to patterns of long-range entanglement. In recent years, it was
shown that in 1+1D bosonic systems there is no nontrivial topological order,
while in 2+1D bosonic systems the topological orders are classified by a pair:
a modular tensor category and a chiral central charge.  In this paper, we
propose a partial classification of topological orders for 3+1D bosonic
systems: If all the point-like excitations are bosons, then such topological
orders are classified by unitary pointed fusion 2-categories, which are
one-to-one labeled by a finite group $G$ and its group 4-cocycle $\omega_4 \in
\cH^4[G;U(1)]$ up to group automorphisms. Furthermore, all such 3+1D
topological orders can be realized by Dijkgraaf-Witten gauge theories.

\end{abstract}

\maketitle

{\small \setcounter{tocdepth}{2} \tableofcontents }

\section{Introduction}

In history, we have completely classified some large class of matter states
only for a few times.  The first time is the classification of all spontaneous
symmetry breaking orders \cite{L3726,LanL58}. We find that all symmetry
breaking orders can be described by a pair: 
\begin{align}
(G_\Psi\subset G_H), 
\end{align}
where  $G_H$ is the symmetry group of the system and $G_\Psi$, a
subgroup of $G_H$, is the symmetry group of the ground state. 

The second time is the classification of all 1-dimensional gapped quantum
phases.  We find that 1-dimensional gapped quantum phases  with on-site
symmetry $G_H$ can be classified by a triple (even for strongly interacting
bosons/fermions) \cite{CGW1107,SPC1139}:
\begin{align}
[G_\Psi\subset G_H;\ \text{pRep}(G_\Psi)], 
\end{align}
where pRep$(G_\Psi)$ is a \emph{projective representation} of
$G_\Psi$ \cite{PBTO0909.4059}.  We see that all the 1-dimensional gapped
quantum phases are described by symmetry breaking plus an addition structure
described by $\text{pRep}(G_\Psi)$. The additional structure is the
symmetry-protect topological (SPT) order\cite{GW0931}.

The third time is the classification of 2-dimensional gapped quantum phases. In
the absence of any symmetry, a gapped phase may have a nontrivial
topological order \cite{Wrig,WNtop,KW9327}. We find that all 2+1D bosonic
topological orders are classified by a pair:\cite{KW9327,RSW0777,W150605768}
\begin{align}
(\text{MTC},c), 
\end{align}
where MTC is a unitary \emph{modular tensor category} and $c$ is  the chiral
central charge of the edge states. Physically, the tensor category theory MTC
is just a theory that describes the fusion and the braiding of  quasiparticles,
which correspond to fractional/non-abelian statistics. Modular means that every
nontrivial quasiparticle has a nontrivial mutual statistics with some
quasiparticles.

For fermion systems, the classification is
different: 2+1D fermionic topological orders are classified by a
triple:\cite{LW150704673}
\begin{align}
[\sRep (Z_2^f) \subset \text{BFC};\ c], 
\end{align}
where $\sRep (Z_2^f)$ is the \emph{symmetric fusion category} (SFC).  In
general, a SFC describes particles with trivial mutual statistics with each
others.  In the SFC $\sRep (Z_2^f)$, the particles carry the representations of
the fermion-number-parity symmetry $Z_2^f$ where the nontrivial representation
is assigned Fermi statistics.  So $\sRep (Z_2^f)$ just describes the underlying
fermions that form the fermionic systems.  Also, BFC is a unitary \emph{braided
fusion category} that describes all the quasiparticles (the anyons and the
underlying fermions). Those quasiparticles all have trivial mutual statistics
with the underlying fermions in $\sRep (Z_2^f)$. We also require BFC to have
minimal modular extensions.  

In the presence of unitary finite on-site symmetry $G_H$, all the 2+1D
gapped bosonic phases are classified by \cite{BBC1440,LKW1602.05946} 
\begin{align}
[G_\Psi\subset G_H;\ 
\Rep (G_\Psi) \subset \text{BFC} \subset \text{MTC};\ c],
\end{align}
where $\Rep (G_\Psi)$ is a SFC that describes bosonic quasiparticles
with trivial mutual statistics that carry the representations of
$G_\Psi$, BFC describes all the quasiparticles that have trivial mutual
statistics with all the quasiparticles in $\Rep (G_\Psi)$, and MTC is a
minimal modular extension of the BFC.  Physically speaking, MTC include both
quasiparticles (described by BFC) and symmetry-twist defects (added by hand), such
that every nontrivial quasiparticle/defect has a nontrivial mutual statistics
with some quasiparticles/defects, and every defect has a nontrivial mutual
statistics with some quasiparticles in $\Rep (G_\Psi)$.  Such a
classification includes symmetry breaking orders (described by
$G_\Psi\subset G_H$), as well as the SPT orders \cite{CGL1314,CGL1204}
[described by $\Rep (G_\Psi) = \text{BFC} \subset \text{MTC},\ c$] and
topological orders [described by $\Rep (1) \subset \text{BFC} = \text{MTC},\
c$].

For fermion systems with unitary finite on-site symmetry $G_H^f$, we have a
very similar classification: all such 2+1D gapped fermionic phases are
classified by \cite{LKW1602.05946} 
\begin{align}
[G_\Psi^f\subset G_H^f;\
\sRep (G_\Psi^f) \subset \text{BFC} \subset \text{MTC};\ c], 
\end{align}
where $\sRep (G_\Psi^f)$ describes quasiparticles with trivial mutual
statistics that carry the representations of $G_\Psi^f$ where some
representations are assigned Fermi statistics.

After those fairly complete classification results in 1+1D and 2+1D, in this
paper, we are going to study the classification of 3+1D topological orders.  We
will only deal with the simpler case, the 3+1D topological orders for bosonic
systems.  3+1D bosonic topological orders are gapped quantum liquids
\cite{ZW1490,SM1403} without any symmetry.  Note that there are gapped
non-liquid states in 3+1D, such as stacked fractional-quantum Hall layers
\cite{ZW1490}, and fractal/fracton topological states
\cite{Cc0404182,BT10064871,VF160304442,MH170100747,V170100762,HH170302973,SK170403870}
which include Haah's model \cite{H11011962}.  Unlike the non-liquid states, the
gapped quantum liquids have point-like excitations and string-like excitations,
that can move in all directions and have nontrivial braidings among them.  But
the statistics for point-like excitations alone is simpler than in 2+1D; they
are bosons or fermions with trivial mutual statistics. In other words, the
point-like excitations in a 3+1D topological order are described fully by a
SFC.

If the point-like excitations are all bosons, the corresponding  SFC will
always have the form $\Rep(G)$ for some finite group $G$.  In other words, the
point-like excitations of a 3+1D topological order can always be viewed as
carrying irreducible representations of the group.\footnote{In this paper, we
will also call those pointed-like excited states carrying reducible group
representations as a (composite) point-like excitation; they in general have
accidental degeneracy.  This is unconventional in physics literature, but
beneficial when we discuss the universal properties, such as fusion, of topological orders. The
composite (accidentally degenerate) excitations also play an important role.}
They exactly behave like the quasiparticle excitations above a product state
with $G$ symmetry.  This is a quite amazing result: a 3+1D topological order
whose quasiparticles are all bosonic  is always related to a finite group
$G$.\footnote{This is true even for the general cases when some quasiparticles
are fermions.} We will refer to such (point-like-excitations-are-)all-boson
topological order as AB topological order.

One may naturally wonder if a 3+1D AB topological order is always described by
a $G$-gauge theory, since the point-like excitations in a $G$-gauge theory are
indeed described by $\Rep(G)$.  In fact, the above statement is not true.
There are 3+1D topological orders arising from the Dijkgraaf-Witten gauge
theory \cite{DW9093}, whose point-like excitations are also described by
$\Rep(G)$.  So we cannot say that all 3+1D topological order with $\Rep(G)$
point-like excitations are described by the usual $G$-gauge theory.  But, do we
have something even more general than Dijkgraaf-Witten gauge theory that also
produce $\Rep(G)$ point-like excitations?  In this paper, we like to show that
there is nothing more general: 
\myfrm{All 3+1D topological orders, whose point-like excitations are all
bosons, are classified by a finite group $G$ and its group 4-cocycle $\om_4 \in
\cH^4[G;U(1)]$, up to group automorphisms.} In this paper, ``classified''
always means a correspondence in a one-to-one fashion.  Furthermore, \myfrm{All
3+1D AB topological orders can be realized by Dijkgraaf-Witten gauge theory
with a finite gauge group.}

The above result is obtained by condensing all the point-like excitations in a
3+1D topological order $\sC^4$ to form a new topological order $\sD^4$ (which
is possible when all the point-like excitations are bosons), and argue that
\begin{enumerate}
\item The new phase $\sD^4$ is a trivial phase. Therefore, $\sC^4$ has a 2+1D
gapped boundary $\cM^3$ induced by such condensation, which carries only
string-like excitations.

\item  The above string-like excitations on the boundary are labeled by the
elements of a finite group $G$, and their fusion rule is given by the group
multiplication. It is the same group whose representations are carried by
the point-like excitations in the bulk.

\item The string-only boundary $\cM^3$ form a unitary pointed fusion 2-category
whose only nontrivial level are the objects. The different pointed fusion
2-categories are classified by a finite group $G$ and its 4-cocycle $\omega_4$
in $H^4(G,U(1))$, up to group automorphisms.

\item The bulk topological order $\sC^4$ is the center of the fusion 2-category
$\cM^3$: $\sC^4 = Z(\cM^3)$ \cite{KW1458,KW170200673}, which is a
Dijkgraaf-Witten gauge theory with $(G,\omega_4)$.  Furthermore, each bulk
topological order $\sC^4$ corresponds to a unique unitary pointed fusion
2-category $\cM^3$.

\end{enumerate}
In the following, we will discuss some general properties of 3+1D topological
orders.  Then we will show the main result of the paper following the above
four steps.

\section{Excitations in topologically ordered state}

\subsection{Point-like excitations}

\subsubsection{Use trap Hamiltonian to define excitations}

Consider a bosonic system defined by a local gapped Hamiltonian $H_0$ in $d$
dimensional space $M^d$ without boundary.  A collection of quasiparticle
excitations labeled by $p_i$ and located at $\v x_i$ can be produced as
\emph{gapped} ground states of $H_0+\sum_i \del H_{p_i}$ where $\del H_{p_i}$
is non-zero only near $\v x_i$.  By choosing different $\del H_{p_i}$'s we can
create (or trap) all kinds of point-like excitations.  The gapped ground states
of $H_0+\sum_i \del H_{p_i}$ may have a degeneracy $D( M^d;p_1,p_2,\cdots )$
which depends on the quasiparticle types $p_1,p_2,\cdots$ and the topology of
the space $M^d$. The degeneracy is not exact, but becomes exact in the large
space and large particle separation limit.  We will use
$\cV(M^d;p_1,p_2,\cdots)$ to denote the space of the degenerate ground states,
which will also be called fusion space.
If the Hamiltonian $H_0+\sum_i \del H_{p_i}$ is not gapped, we will say $D(
M^d;p_1,p_2,\cdots)=0$ ({i.e.,}\ $\mathcal V(M^d;p_1,p_2,\cdots )$ has zero
dimension).  If $H_0+\sum_i \del H_{p_i}$ is gapped, but if $\del H_{p_i}$ also
creates quasiparticles away from $\v x_i$'s (indicated by the bump in the
energy density away from $\v x_i$'s), we will also say $D( M^d;p_1,p_2,\cdots
)=0$.  (In this case quasiparticles at $\v x_i$'s do not fuse to trivial
quasiparticles.) So, if $D( M^d;p_1,p_2,\cdots)>0$, $\del H_{p_i}$ only
creates/traps quasiparticles at $\v x_i$'s.

For topologically
ordered state with no spontaneous symmetry breaking, the fusion space on
$d$-dimensional sphere $M^d=S^d$ with no
particles $\cV(S^d)$ is always one dimensional.  Thus in the presence of
point-like excitations, dimension of the fusion space, $\cV( S^d;p_1,p_2,\cdots
)$ represents the total number of internal degrees of freedom for the
quasiparticles $p_1,p_2,\cdots )$.  To obtain the  number of internal degrees
of freedom for type-$p_i$ quasiparticle, we consider the dimension $D(
S^d;p_i,p_i,\cdots, p_i)$ of the fusion space on $n$ type-$p_i$ particles on
$S^d$.  In large $n$ limit $D( S^d;p_i,p_i,\cdots, p_i)$ has a form
\begin{align}
 \ln D( S^d;p_i,p_i,\cdots, p_i) =n (\ln d_{p_i} +o(1/n) ).
\end{align}
Here $d_{p_i}$ is called the quantum dimension of the type-$p_i$ particle,
which describe the internal degrees of freedom the particle.  For example, a
spin-0 particle has a quantum dimension $d=1$, while a spin-1  particle has a
quantum dimension $d=3$.

\subsubsection{Simple type and composite type} \label{types}

Two excitations $p$ (trapped by $\Del H_p$) and $p'$ (trapped by $\Del H_{p'}$)
are said to have the same type if the corresponding fusion spaces
$\cV(S^d;p,p_1,\cdots )$ and $\cV(S^d;p',p_1,\cdots )$ can smoothly deform into
each other as we change the trap Hamiltonian from $\Del H_p$ to $\Del H_{p'}$.
Two excitations $p$ and $p'$ are of the same type iff they only differ by some
local operators.  If an exaction can be created by local operators from the
ground state, the excitation will be said to have a trivial type, and denoted
as $\one$.

Even after quotient out the local excitations of trivial type, topological
quasiparticle type still have two kinds: \emph{simple type} and \emph{composite
type}: If the ground-state degenerate subspace $\cV(M^d;p,q,\cdots )$ cannot
not be splitted  by any small local perturbations near $\Del H_p$, then the
particle $p$ is said to be simple. Otherwise, the  particle type $p$ is said to
be composite.  

When $p$ is composite, the fusion space 
$\cV(M^d;p,q,\cdots)$ has a direct sum decomposition (after splitting by a
generic perturbation of $\Del H_p$):
\begin{align}
&\quad\ \cV(M^d;p,q,\cdots)
\nonumber \\
 &=
\cV(M^d;p_1,q,\cdots)\oplus \cV(M^d;p_2,q,\cdots)
\nonumber \\
& \ \ \ \ \ \ \
\oplus \cV(M^d;p_3,q\cdots)\oplus \cdots
\end{align}
 where $p_1$, $p_2$, $p_3$, {\it etc.} are simple types.  The above
decomposition allows us to denote the composite type $i_1$ as
\begin{align}
 p=p_1\oplus p_2\oplus p_3\oplus
\cdots. 
\end{align}

\subsubsection{Fusion of point-like excitations}

When we fuse two simple types of topological particles $p_1$
and $p_2$ together, it may become a topological particle of a composite type:
\begin{align} 
p_1\otimes p_2=q=p_3'\oplus p_3'' \oplus \cdots.
\end{align}
Here, we will use an integer tensor $N^{p_1p_2}_{p_3}$ to describe the
quasiparticle fusion, where $p_i$ label simple types:  
\begin{align}
p_1\otimes p_2=\bigoplus_{p_3} N^{p_1p_2}_{p_3} p_3 . 
\end{align}
Such an integer tensor $N^{p_1p_2}_{p_3}$ is referred as the
fusion coefficients of the topological order, which is a universal property of
the topologically ordered state.

The internal degrees of freedom (\ie the quantum dimension $d_p$) for the
type-$p$ simple particle can be calculated directly from $N^{p_1p_2}_{p_3}$.
In fact $d_p$ is the largest eigenvalue of the matrix $N_p$, whose elements are
$(N_{p})_{p_2p_1} = N^{pp_1}_{p_2}$. 

\subsection{String-like excitations}

Similarly, we can also use \emph{gapped} trap Hamiltonians $H_0+\sum_s \Del
H_s$ to define string-like excitations, where $\Del H_s$ is no zero only near a
loop.  The ground state subspace of $H_0+\sum_s \Del H_s$ is called the fusion
space of strings $\cV(M^d,s,t,\cdots)$.  If the fusion spaces
$\cV(M^d,s,t,\cdots)$ and $\cV(M^d,s',t,\cdots)$ can smoothly deform into each
other, we say the strings $s$ and $s'$ are of the same type.

If the ground-state degenerate subspace $\cV(M^d;s,t,\cdots )$ cannot not be
split by any small \emph{non-local} perturbations along the string $s$,
then the string $s$ is said to be simple. Otherwise, the string $s$ is said to
be composite.  We stress that here we allow non-local perturbations along the
string $s$.  In other word, any degrees of freedom near the string can interact
no matter how far are them.  But the interactions do not involve degrees of
freedom far away from the string.  The non-local perturbations is necessary. If
we used local  perturbations to define string types, we would have too many
string types that are not related to the topological orders in the ground
state.

The fusion of the string loops is also described by integer tensors
\begin{align}
 s_1\otimes s_2 = \bigoplus_{s_3} M^{s_1s_2}_{s_3} s_3.
\end{align}
The string loops can also shrink and become point-like excitations
\begin{align}
 s_i \to \bigoplus_{p_j} M^{s_i}_{p_j} p_j .
\end{align}
We like to conjecture that \myfrm{If $M^{s}_{\one} >1$, then the string $s$ is
not simple (\ie $s$ is a direct sum of several strings).} If a simple string
satisfy $M^{s}_{\one} =1$, we say $s$ is a \emph{pure} simple string.

\subsection{The on-string excitations are always gappable}

The strings, as 1D extended objects, may carry excitations that travel along
them. Those excitations can some times be gapless.  In the following, we
like to argue that, if the types of point-like and string-like excitations are
finite, those on-string excitations can always be gapped by adding
proper interactions.

We do this by contradiction. Assuming that on-string excitations cannot be
gapped by interactions, based on what we know about 1+1D system, there are only two situations.\\
\begin{enumerate}
\item When the on-string excitations are chiral with a non-zero chiral central
charge $c$.   But in this case, when fusing $n$ string together, the new string
will have on-string excitations with chiral central charge $nc$.  This means
that the fusion will produce infinite types of strings.
The finite string-type assumption excludes the possibility of on-string excitations with a non-zero chiral
central charge.

\item When the on-string excitations are described by the edge of certain
fractional-quantum Hall states (which have chiral central charge $c=0$)
\cite{L1309}.  In this case, those on-string excitations have a gravitational
anomaly described by a non-invertible 2+1D topological order
\cite{W1313,KW1458} In this case, the open membrane operator that creates such a
string on its boundary must creates a non-invertible 2+1D topological order on
the membrane.  Multiplying membrane operators corresponds to stacking  2+1D
topological orders together, and stacking non-invertible topological order can
never produce a trivial topological order \cite{KW1458}.  Thus  fusing $n$
string together will always produce a new non-trivial string.  Again, finite
string-type assumption exclude this possibility.

\end{enumerate}

We like to remark that ungappable strings can appear in gapped non-liquid
states \cite{ZW1490,SM1403}, such as the 3+1D gapped states obtained by stack
fractional quantum Hall layers.  But in that case, the strings are not mobile
in all the directions.  It appears that the liquid assumption of topological
order \cite{ZW1490,SM1403} makes all strings gappable in 3+1D.  In the rest of
this paper, we will always assume the on-string excitations to be gapped.

\section{Some general properties of
3+1D topological orders of boson systems}

\subsection{The group structure in 3+1D topological order}
\label{grpstrct}

We note that the point-like excitations in 3+1D topological orders are
described by a symmetric fusion category (SFC). Physically, a SFC is just a
collection of particles which are all bosons or fermions with trivial mutual
statistics.

Mathematically, it has been shown that a SFC
must be either $\Rep(G)$ (a braided fusion category (BFC) formed by the
representations of $G$ with all the irreducible representations being assigned
Bose statistics) or $\sRep(G)$ (a braided fusion category (BFC) formed by the
representations of $G$ with some of the irreducible representations being
assigned Bose statistics while other irreducible representations being assigned
Fermi statistics) for some group $G$. 

The above implies that \myfrm{Each 3+1D topological order is
associated with a group $G$, where the point-like excitations (particles) are described by $\Rep(G)$ or
$\sRep(G)$.} In this paper, we will use this fact heavily to gain a systematic
understanding of 3+1D topological orders.  (In fact, each higher dimensional
topological order is also related to a group in the same fashion.) In some
sense, 3+1D topological orders can all be viewed as gauge theories with some
old or new twists.


However, those point-like excitations have trivial mutual statistics among
them.  One cannot use the point-like excitations to detect other  point-like
excitations by remote operations.  In general, we believe \cite{L1309,KW1458} \myfrm{ \textbf{The
principle of remote detectability:} In an anomaly-free topological order, every 
topological excitation can be detected by other topological excitations via
some remote operations.  If every topological excitation can be detected by
other topological excitations via some remote operations, then the topological
order is anomaly-free.  } Here ``anomaly-free'' means realizable by a local
bosonic model on lattice \cite{W1313}. Thus the  remote detectability condition
is also the anomaly-free condition.

The above implies that an anomaly-free (\ie realizable) 3+1D topological order
must contain string-like topological excitations, so that every  point-like
topological excitation can be detected by some string-like topological
excitations via remote braiding, and every string-like topological excitation
can be detected by some point-like and/or string-like topological excitations
via remote braiding.  We see that the properties of string-like topological
excitations are determined by the point-like topological excitations (\ie
$\Rep(G)$ or $\sRep(G)$) to a certain degree.

\subsection{Dimension reduction of topological orders}

\label{dimred}

\begin{figure}[b] 
\centering \includegraphics[scale=0.6]{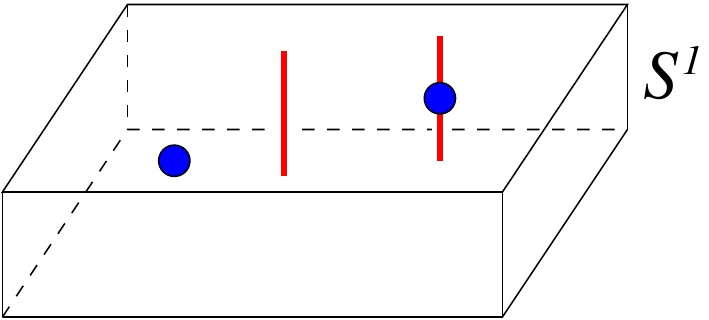} 
\caption{(Color online)
The dimension reduction of 3D space $M^2\times S^1$ to 2D space $M^2$. The top
and the bottom surfaces are identified and the vertical direction is the
compactified $S^1$ direction.  A 3D point-like excitation (the blue dot)
becomes an anyon particle in 2D.  A 3D string-like excitation wrapping around
$S^1$ (the red line) also becomes an anyon particle in 2D.
}
\label{3D2D} 
\end{figure}

To understand better the
relation between the point-like and the string-like excitations, we will
introduce  the dimension reduction in the next section, which turns out to be a
very useful tool in our approach.
We can reduce a $d+1$D topological order $\sC^{d+1}$ on space-time
$M^d\times S^1$ to $d$D topological orders on space-time $M^d$ by making the
circle $S^1$ small (see Fig. \ref{3D2D}).  In this limit, the $d+1$D
topological order $\sC^{d+1}$ can be viewed as several $d$D topological orders
$\sC^{d}_i$, $i=1,2,\cdots,N^\text{sec}_1$ which happen to have degenerate
ground state energy.  We denote such a dimensional reduction process by
\begin{equation}
\label{C3C2}
\sC^{d+1} = \bigoplus_{i=1}^{N^\text{sec}_1} \sC^{d}_i,
\end{equation}
where $N^\text{sec}_1$ is the number of sectors 
produced by the dimensional reduction.

\begin{table}[t]
\caption{
The dimension reduction $M^3\times S^1$ to $M^3$ of the 3+1D $S_3$-gauge theory
$\sC^4_{S_3}$, where $G_\chi$'s are $S_3,Z_2,Z_3$.  The 3+1D point-like
excitations $p_0,p_1,p_2$ becomes the 2+1D point-like excitations.  The 3+1D
loop-like excitations $s_{\chi q}$, when wrapped around the $S_1$, also becomes
the 2+1D point-like excitations.  $\sC^3_{S_3}$ is the untwisted sector where
$\one,A^1,A^2$ correspond to $\Rep(S_3)$.  See \Ref{MW1514} and Appendix
\ref{gauge}.
}
\label{Dred}
\centering
\begin{tabular}{|c|c|c|c|}
\hline
 $\sC^4_{S_3} \to $ & $\sC^3_{S_3}$ & $\sC^3_{Z_2}$ & $\sC^3_{Z_3}$   \\
\hline
Symmetry Breaking	& $S_3\rightarrow S_3$ & $S_3\rightarrow \mathbb Z_2$	& $S_3\rightarrow \mathbb Z_3$\\
$ p_0\rightarrow$ &	$\bm 1$	& $\bm 1$	& $\bm 1$	\\
$p_1\rightarrow$	  & $A^1$	& $e$	& $\bm 1$	\\
$p_2\rightarrow$	  &	$A^2$	& $\bm 1\oplus e$	& $e_1\oplus e_2$	\\
$s_{20}\rightarrow$	  	  &	$B$		&	m	&	-\\
$s_{21}\rightarrow$	  &	$B^1$	&	em	&	-\\
$s_{30}\rightarrow$	  	  &	$C$		& -	& $m_1\oplus m_2$	\\
$s_{31}\rightarrow$	  &	$C^1$	& -	& $e_1m_1\oplus e_1m_2$	\\
$s_{32}\rightarrow$	  &	$C^2$	& -	& $e_2m_1\oplus e_2m_2$	\\
\hline
\end{tabular}
\end{table}

For example, let us use $\sC^{d+1}_G$ to denote the $d+1$D topological order
described by the gauge theory with the finite gauge group $G$. We find that,
for $d\geq 3$ (see Table \ref{Dred}) \cite{MW1514},
\begin{align}
\label{G3G2}
\sC_G^{d+1} = \bigoplus_{\chi}\sC_{G_\chi}^{d} 
\end{align}
where $\bigoplus_{\chi}$ sums over all different conjugacy classes $\chi$ of
$G$, and $G_\chi$ is a subgroup of $G$ formed by all the elements that commute
with an element in $\chi$.  In fact, each dimension reduced $d$D topological
order, $\sC_{G_\chi}^{d}$, is produced by threading a $G$-gauge flux described
by the conjugacy classes $\chi$ through the $S^1$ in the space-time $M^d\times
S^1$.  The $\chi$-flux breaks the gauge symmetry $G$ down to $G_\chi$.  Thus
the corresponding $d$D topological order is a $G_\chi$-gauge theory.

For Dijkgraaf-Witten theories (gauge theories twisted by group-cocycles of the
gauge group), the dimension reduction have a form \cite{WW1404.7854}
\begin{align}
\label{G3G2w}
\sC_{G,\om_{d+1}^G}^{d+1} = \bigoplus_{\chi}\sC_{G_\chi,\om_d^{G_\chi}(\chi)}^{d} ,
\end{align}
where $\om_{d+1}^G$ is a $(d+1)$-group-cocycle $\om_{d+1}^G \in
\cH^{d+1}(G;U(1))$, $\om_d^{G_\chi}(\chi)$ is a $d$-group-cocycle
$\om_d^{G_\chi}(\chi) \in \cH^d(G_\chi;U(1))$, and $\sC_{G,\om_{d+1}^G}^{d+1}$
is the topological order described by Dijkgraaf-Witten theory with gauge group
$G$ and cocycle twist $\om_{d+1}^G$.


To understand the number of sectors $N^\text{sec}_1$ in the dimension
reduction, we note that the different sectors come from the different holonomy
of moving point-like excitations around the $S^1$ (see Fig. \ref{3D2D}).  For gauge theory, this so
called holonomy comes from the gauge flux going through the compactified
$S^1$.  For more general topological orders, this holonomy comes from
threading co-dimension 2 topological excitations through the $S^1$.  

From this picture, we see that the number of sectors
$N^\text{sec}_1$ is bounded by the number of types of the co-dimension 2 pure
topological excitations.  Also, if two co-dimension 2 topological excitations
cannot be distinguished by their braiding with point-like excitations, then
threading them through the $S^1$  will not produce different sectors.  Thus, the
number of sectors $N^\text{sec}_1$ is the number of the classes of
co-dimension 2 topological excitations that can be distinguished by the
braiding with the point-like excitations.  In particular, in 3+1D, the number
of the sectors $N^\text{sec}_1$ is the number of the classes of string-like
topological excitations that can be distinguished by the braiding with the
point-like excitations.

From the above discussion, we also see that the dimension reduction always
contain a sector where we do not thread any nontrivial string through $S^1$ and
the holonomy of moving any point-like excitations around the $S^1$ is trivial.
Such a sector will be called the untwisted sector.  Also, if a topological
order has no nontrivial point-like excitation, then its dimension reduction
contains only one sector -- the untwisted sector.

In the untwisted sector, there are three kinds of anyons.  The first kind of
anyons correspond to the 3+1D point-like excitations.  The second kind of
anyons correspond to the 3+1D pure string-like excitations wrapping around the
compactified $S^1$.  The third kind of anyons are bound states of the first two
kinds (see Fig. \ref{3D2D}).

\begin{figure}[tb] 
\centering \includegraphics[scale=0.4]{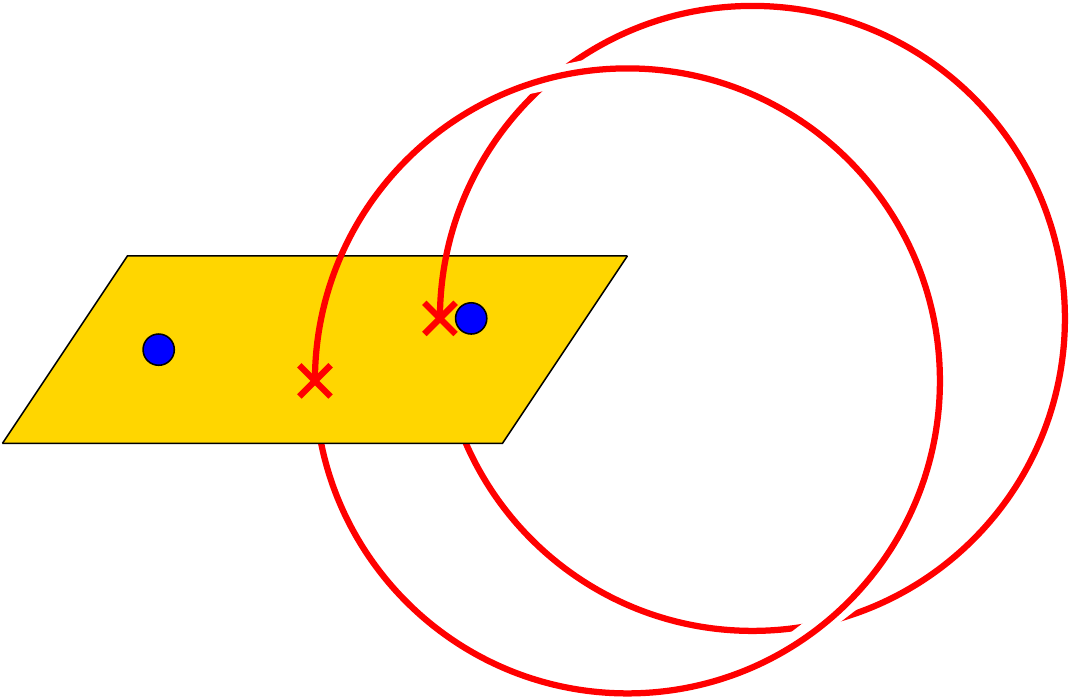} 
\caption{(Color online)
The untwisted sector in the dimension reduction can be
realized directly on a 2D sub-manifold in 3D space without compactification.
}
\label{3D2Dsub} 
\end{figure}

We like to point out that the untwisted sector in the dimension reduction can
even be realized directly in 3D space without compactification. Consider a 2D
sub-manifold in the 3D space (see Fig. \ref{3D2Dsub}), and put the 3D point-like
excitations on the 2D sub-manifold.  We can have a loop of string across the 2D
sub-manifold which can be viewed as an effective point-like excitation on the 2D
sub-manifold. We can also have a bound state of the above two types of effective
point-like excitations on the 2D sub-manifold.  Those effective point-like
excitations on the 2D sub-manifold can fuse and braid just like the anyons in
2+1D.  The principle of remote detectability requires those effective point-like
excitations to form a MTC.  When we perform dimension reduction, the above MTC
becomes the untwisted sector of the dimension reduced 2+1D topological order.
We like to mention that the dimension reduction introduce new types of the
perturbations that may not be local from 3+1D point of view.  But those new
perturbations are local in the dimension reduced 2+1D theory.  MTC is very
rigid which cannot be changed by any 2+1D  local perturbations.  This is why
the untwisted sector is still described by the same MTC that describes the
effective point-like excitations on the 2D sub-manifold.

This way, we show that \myfrm{The distinct (simple) point-like excitations in
the 3+1D topological order become the distinct anyons (\ie the simple objects)
in the dimension reduced 2+1D topological order that corresponds to the
untwisted sector.} Table \ref{Dred} describes the dimension reduction of a 3+1D
topological order $\sC^4_{S_3}$ described by $S_3$-gauge theory:
\begin{align}
 \sC^4_{S_3} \to \sC^3_{S_3} \oplus\sC^3_{Z_2} \oplus\sC^3_{Z_3} .
\end{align}
The 2+1D topological order $\sC^3_{S_3}$ is the untwisted sector.  The three
types of particles in 3+1D $\sC^4_{S_3}$, $p_0,p_1,p_2$, that form a SFC
$\Rep(S_3)$ becomes three types of particles in the untwisted sector
$\sC^3_{S_3}$, $\one,A^1,A^2$, that also form a $\Rep(S_3)$.  We also see
that in other sectors, the distinct point-like excitations in 3+1D may not be
reduced to distinct simple objects in the dimension reduced 2+1D topological
orders.

Since the dimension reduced 2+1D topological orders must be anomaly-free, they
must be described by modular tensor category.  Since the untwisted sector
always contains $\Rep(G)$, we conclude that \myfrm{The untwisted sector of a
dimension reduced 3+1D topological order is a modular extension of $\Rep(G)$.}
In next section, we will show that such a modular extension must be a minimal
one.

\subsection{Untwisted sector of dimension reduction is the Drinfeld center
of $\sE$}
\label{untw}

In the following we will show a stronger result, for the untwisted sector.  
Given a 3+1D bosonic topological order, let the symmetric fusion category
formed by the point-like excitations be $\sE$, $\sE=\Rep(G)$ or
$\sE=\sRep(G^f)$.  \myfrm{The untwisted sector $\sC^3_\text{untw}$ of dimension
reduction of a 3+1D topological orders must be the 2+1D topological order
described by Drinfeld center of $\sE$: $\sC^3_\text{untw}=Z(\sE)$.} Note that
Drinfeld center $Z(\sE)$ is the minimal modular extension of $\sE$.

First, let us recall the definition of Drinfeld center. The Drinfeld center
$Z(\cA)$ of a fusion category $\cA$, is a braided fusion category, whose
objects are pairs $(A,b_{A,-})$, where $A$ is an object in $\cA$, $b_{A,-}$ is
a set of isomorphisms $b_{A,X}:A\otimes X\cong X\otimes A,\forall X\in \cA$,
satisfying natural conditions. $b_{A,X}$ is called a half braiding. Morphisms
between the pairs $(A,b_{A,-}),(B,b_{B,-})$ is a subset of morphisms between
$A,B$, such that they commute with the half braidings $b_{A,-},b_{B,-}$. The
fusion and braiding of pairs is given by
\begin{align}
  (A,b_{A,-})\otimes (B,b_{B,-})&=(A\otimes B,
  (b_{A,-}\otimes\id_B)(\id_A\otimes b_{B,-}),\nonumber\\
  c_{(A,b_{A,-}), (B,b_{B,-})}&=b_{A,B}.
\end{align}
In other words, to half-braid $A\otimes B$, one just half-braids $B$ and $A$
successively, and the braiding between pairs is nothing but the half braiding.

$\sC^3_\text{untw}=Z(\sE)$ is the consequence that the strings in the untwisted
sectors are in fact shrinkable. From the effective theory point of view, we can
shrink a string $s$ (including bound states of particles with strings, in
particular, point-like excitations viewed as bound states with the trivial
string) to a point-like excitation $p^\text{shr}_s$ in $\sE$
\begin{align}
\label{stop}
  s\to p^\text{shr}_s=p_1\oplus p_2\oplus\dots, \quad p_1,p_2,\dots\in\sE
\end{align}
So if we only consider fusion, the particles $s,\ p$ in the dimension reduced
untwisted sector $\sC^3_\text{untw}$ can all be viewed as the particles in
$\sE$, regardless if they come from the 3D particles or 3D strings. In
particular, the particles from the 3+1D strings $s$ can be viewed as composite
particles in $\sE$ (see \eqn{stop}). 
To obtain the Drinfeld center we
need to introduce braiding among those particles $p$'s and $s$'s. 

In the untwisted sector, the braiding between strings $s,s'$, denoted by
$c_{s,s'}$, requires string $s'$
moving through string $s$, which prohibits shrinking string $s$.
However,
there is no harm to consider the shrinking if we focus on only the initial and
end states of the braiding process. 

\begin{figure}[tb]
  \centering
  \includegraphics[scale=0.5]{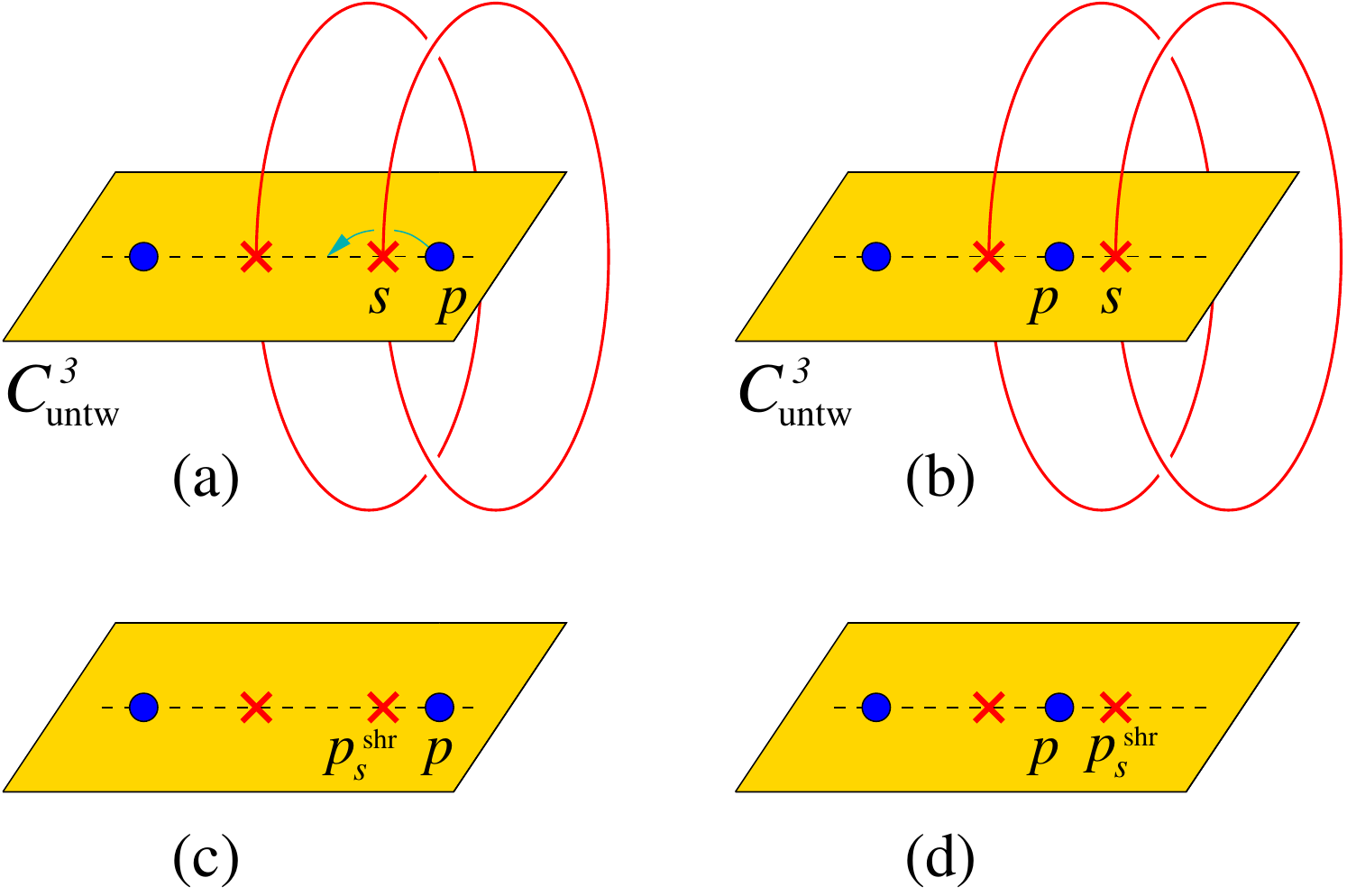}
  \caption{(Color online)
  From (a) to (b) is the braiding $c_{s,p}$ in the untwisted sector.
(c)(d) are obtained from (a)(b) by shrinking strings. Shrinking thus induces a
``half-braiding'' isomorphism $c^\text{shr}_{s,p}$ from (c) to (d).}
  \label{fig:hbraid}
\end{figure}

In particular, the braiding between a string $s$ and a particle $p$, induces an
isomorphism between the initial and end states where the string $s$ is shrunk
(see Fig.~\ref{fig:hbraid})
\begin{align}
  c^\text{shr}_{s,p}: p^\text{shr}_s\otimes p \cong p \otimes p^\text{shr}_s
\end{align}
which is automatically a half-braiding on the particle $p^\text{shr}_s$.
Thus, $(p^\text{shr}_s,c^\text{shr}_{s,-})$, by definition, is an object in
the Drinfeld center $Z(\sE)$.
Shrinking induces a functor
\begin{align}
  \sC^3_\text{untw}&\to Z(\sE)\nonumber\\
  s&\mapsto (p^\text{shr}_s,c^\text{shr}_{s,-})
\end{align}
which is obviously monoidal and braided, \ie, preserves fusion and braiding. It is also fully faithful, namely
bijective on the morphisms. Physically this means that the local 
operators on both sides are the same.
On the left side, morphisms on a string $s$ are operators acting on near (local to)
the string $s$; on the right side, morphisms in the Drinfeld center are
morphisms
on the particle $p^\text{shr}_s$ which commute with the half braiding
$c^\text{shr}_{s,-}$. From the shrinking picture, morphisms on $p^\text{shr}_s$ can be
viewed as the operators acting on both near the string $s$ and the interior
of the string (namely on a disk $D^2$). But in order to commute with $c_{s,p}$
for all $p$, which can be represented by string operators for all $p$ going through the
interior of the string $s$ (this includes all possible string operators,
because string operators for all particles form a basis), we can take only the operators that act trivially on
the interior of the string. Therefore, morphisms on the right side are also
operators acting on only near the string. This establishes that the functor is
fully faithful, thus a braided monoidal embedding functor; in other words,
$\sC^3_\text{untw}$ can be viewed as a full sub-MTC of $Z(\sE)$. However, $Z(\sE)$ is already a
\emph{minimal} modular extension of $\sE$, which implies that
\begin{align}
  \sC^3_\text{untw}=Z(\sE).
\end{align}

As $Z(\sE)$ is known well, many properties can be easily extracted. For
example, objects in $Z(\sE)$ have the form $(\chi, \rho)$,
where $\chi$ is a conjugacy class, $\rho$ is a representation of the
subgroup that centralizes $\chi$. One then
concludes
 \myfrm{1. A loop-like excitation in a 3+1D topological
order always has an integer quantum dimension, which is $|\chi|\dim \rho$.\\
2. Pure strings ($\rho$ trivial) always correspond to conjugacy classes of the group.}

%

We also see that
{a 3+1D bosonic topological order is similar to a gauge theory of a finite group $G$. The following properties are the same:\\
1) the quantum dimensions of point-like and string-like\\
\hspace*{3mm} excitations.\\
2) the fusion rule of those  excitatios. \\
3) particle-loop and two-loop braidings.}

\section{Condensing all the point-like  excitations to obtain a trivial topological order}

Starting from this section, we are going to show the main result of the paper
via the four steps outlined at the end of introduction.  First, we like to show
that condensing all the point-like  excitations in 3+1D always gives us a
trivial topological order.  To do so, we first like to show the following:

\subsection{There is no 3+1D topological order with only
nontrivial string-like excitations}

\begin{figure}[tb] 
\centering \includegraphics[scale=0.55]{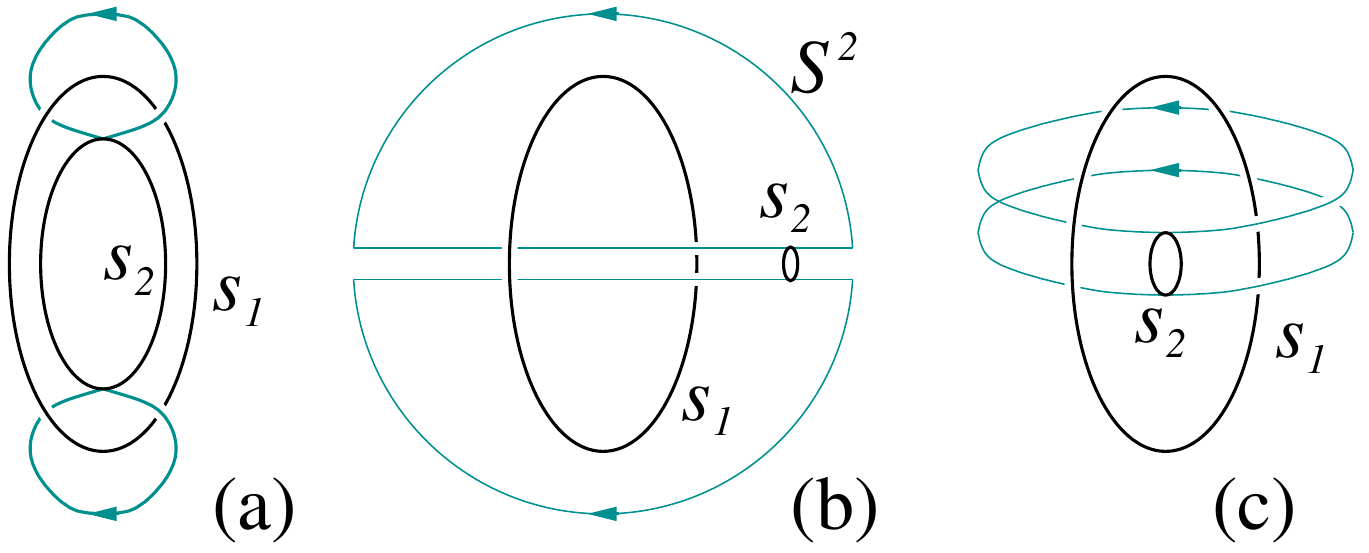} 
\caption{ (Color online)
The braiding path of moving the string $s_2$ around $s_1$.
(a), (b), (c) described the same kind of the braiding paths
that can deform into each other smoothly.
}
\label{looploop} 
\end{figure}

Such a result can be shown using the principle of remote detectability in
Section \ref{grpstrct}.  When there is no nontrivial point-like excitations,
the remote detectability condition requires that a single loop of string can be
remotely detected by braiding other string-like excitations around the loop.
Such a braiding is the two-string braiding described by Fig.  \ref{looploop}a
where a string $s_2$ is braided around a loop $s_1$.

We can also use the dimension reduction picture to show that the anomaly-free
condition requires that the two-string braiding must be nontrivial.  Since
there is no nontrivial point-like excitations, the dimension reduction contains
only the untwisted sector.  In the 2+1D dimension reduced topological order,
all the nontrivial anyons come from the pure strings in the 3+1D topological
order, and correspond to the pure strings wrapping around the compactified
$S^1$.  The 2+1D topological order is anomaly-free and the anyons form a
modular tensor category.  Physically, it means that any nontrivial anyon
must have nontrivial mutual statistics with some anyons. This implies that any
nontrivial pure strings in 3+1D must have nontrivial two-string braiding with
some strings.

Next we like to show that the two-string braiding is always trivial when there
is no nontrivial point-like excitations.  This is because the braiding path of
$s_2$-string around a $s_1$-string in Fig. \ref{looploop}a is a torus wrapped
around the loop $s_1$.  Such a torus can be deformed into a sphere $S^2$ around
$s_1$ with a thin tube going through its center (see Fig. \ref{looploop}b).  If
the total space is a 3-sphere $S^3$, we can deform the sphere $S^2$ into a
small sphere on the other side of $S^3$.  This deforms the  braiding path of
$s_2$ into a thin torus, that describes a small string $s_2$ braiding around
the loop $s_1$ (see Fig.  \ref{looploop}c).  This is like shrinking the string
$s_2$ into a point and let the point braid around the loop $s_1$, Since there
is no nontrivial point-like excitations, the point that represents the small
$s_2$ must have the trivial braiding phase around the loop $s_1$.  This way, we
show that the two strings must have trivial braiding around each other when
there is no nontrivial point-like excitations.  Therefore, \myfrm{The 3+1D
topological orders with only string-like excitations cannot exist (\ie they
must be anomalous).}

\subsection{There is no nontrivial string-like excitations that have trivial
braiding with all  point-like excitations} \label{partstrbraid}

Let us assume that there is a nontrivial string-like excitation $s$, that has
trivial braiding with all  point-like excitations.  If all the point-like
excitations are bosons, then we can condense all the point-like excitations to
obtain a new 3+1D topological order, which will have no nontrivial point-like
excitations. But since the string $s$ has trivial braiding with all point-like
excitations, it can survive the condensation and become a nontrivial
string-like excitation in the new 3+1D topological order.

However, in the last section, we have shown that 3+1D topological orders
with only string-like excitations cannot exist. This contradiction implies that
\myfrm{There is no nontrivial string-like excitations with trivial braiding
with all point-like excitations, if all the point-like excitations are bosons.} 

This result also implies that \myfrm{The untwisted sector of a dimension
reduced 3+1D topological order is a minimal modular extension of $\Rep(G)$.}
This is because, in the untwisted sector, other anyons beside $\Rep(G)$ all
come from strings in 3+1D, which all have nontrivial braiding with the
particles in $\Rep(G)$. This implies the modular extension to be minimal.
However, the above result is weaker than that obtained in Section. \ref{untw}.

\subsection{Condensing all the point-like excitations gives rise to a trivial
3+1D topological order}

When all the point-like excitations are bosons, we can obtain a new
topological order by condensing all the point-like excitations.  The new
topological order has no nontrivial point-like excitations (since they are all
condensed) and has no nontrivial string-like excitations (they are
confined due to the nontrivial braiding with the point-like excitations).  Thus
the new topological order must be an invertible topological order .  But in
3+1D all the invertible topological orders are the trivial one
\cite{K1459,KW1458,F1478}.  Hence condensing all the point-like excitations
gives us a trivial 3+1D topological order.

In a gauge theory, condensing all the point-like excitations corresponds to
condensing all the charged excitations, which breaks all the ``gauge
symmetry''.  This will give us an Anderson-Higgs phase, which is a trivial
phase with no topological order.

\section{String-only boundary of  3+1D topological order}

In this section, we are going to study a particular boundary of 3+1D
topological orders.  We note that there is no gravitational Chern-Simons term
in 3+1D. Thus all 3+1D bosonic topological orders can have a gapped boundary
\cite{KW1458}.  We call such a gapped boundary an anomalous 2+1D
topological order.

From the last section, we see that all 3+1D AB topological orders 
(where all point-like excitations are bosons) can have a gapped boundary
obtained by condensing all the point-like excitations.  For such a boundary,
the anomalous 2+1D topological order on the boundary has no point-like
excitations, and has only string-like excitations.  We will call such a
boundary string-only boundary. Thus \myfrm{All 3+1D AB topological orders  can
have a string-only gapped boundary.}

\subsection{Unitary pointed fusion 2-category}

We will show that the string-only gapped boundary is described by a so called
\emph{unitary pointed fusion 2-category}.  But what is a fusion 2-categories?
In general, a fusion category describes the fusion of codimension-1
excitations, \ie domain-wall excitations.  In 1-dimensional space, the
domain-wall excitations are point-like.  The fusion of those point-like
excitations in 1D space is described by a fusion 1-category (which is also
called fusion category).  In 2-dimensional space, the domain-wall excitations
are string-like.  The fusion of those string-like excitations in 2D space is
described by a fusion 2-category.

\begin{figure}[tb] 
\centering \includegraphics[scale=0.5]{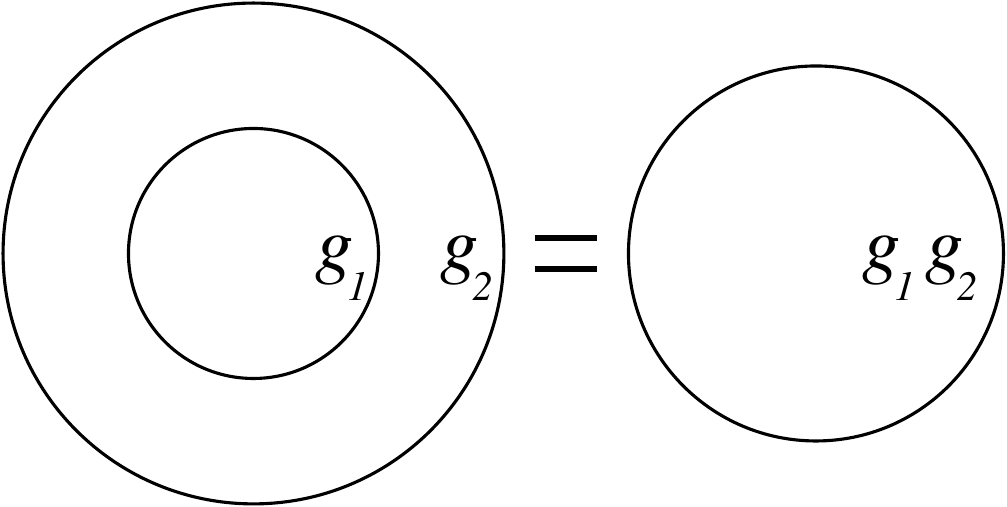} 
\caption{
The fusion of boundary string-like excitations $s^\text{bdry}_{g_1} \otimes
s^\text{bdry}_{g_2} = s^\text{bdry}_{g_1 g_2} $ which can be
abbreviated as $g_1
\otimes g_2 = g_1 g_2 $.
}
\label{strfuse} 
\end{figure}

We like to point out that the fusion 2-categories that describes the
string-only boundary of 3+1D AB topological order are very special: (1)
the string-like excitations on the boundary are labeled by the group elements
of $G$: $s^\text{bdry}_{g},\ g\in G$.  (2) The fusion of the boundary
string-like excitations (see Fig. \ref{strfuse}) is very simple and is given by
the group multiplication
\begin{align}
\label{sssbdry}
 s^\text{bdry}_{g_1} \otimes s^\text{bdry}_{g_2} = s^\text{bdry}_{g_1 g_2} . 
\end{align}
The fusion 2-categories with the above type of fusion rule are called pointed
fusion 2-categories.  Such an amazing result is a consequence of condensing all
the point-like excitations described by $\Rep(G)$ on the boundary.  

One way to show the above result is to consider the untwisted sector of the
dimension reduction, which is a 2+1D topological order.  We have shown that the
untwisted sector is a minimal modular extension of $\Rep(G)$ in Sections
\ref{untw} and \ref{partstrbraid}.  The 2+1D boundary with only strings
corresponds to a 1+1D boundary with only particles of the untwisted sector in
the dimension reduced 2+1D topological order.  Such a 1+1D boundary is obtained
by condensing all the anyons in $\Rep(G)$. The corresponding mathematical
problem has already been solved, see for example \Ref{DGNO09}; we reorganized
the related mathematical results and provided physical interpretations in
\Ref{LKW1602.05946} (Section VI\,D). We find that the particles on such 1+1D
boundary of the untwisted sector are labeled by group elements in $G$ with a
fusion given by group multiplication. Those 1+1D boundary particles correspond
to  the strings on the 2+1D boundary (see Fig. \ref{3D2D}), this allows us to
show \eqn{sssbdry}.  In the next a few sections, we will give a different
argument without using dimension reduction.

\subsection{Tannaka Duality in more explicit language}

Our argument relies heavily on the Tannaka duality, or Tannaka reconstruction
theorem for group representations. It is exactly how we extract the group $G$
from an abstract symmetric fusion category (SFC). A naive example is that for
an abelian group, the tensor product of its irreducible representations, has
exactly the same group structure, which can be viewed as a Fourier
transformation.

In more general cases, one can reconstruct a group $G$ from its representation
category $\Rep(G)$, by the automorphisms of a fiber functor, namely a
functor $F$ from $\Rep(G)$ to the category of vector spaces $\Ve$, that
preserves the fusion and braiding.  We know that the  category of vector spaces
$\Ve$ describes particles in a trivial phase (\ie in a
product state with no symmetry). So one way to physically realize a fiber
functor is by condensing (or other ways such as symmetry breaking) a nontrivial phase to a trivial phase.
With a fiber functor $F$, we have
\begin{align}
  G\cong \Aut(F:\Rep(G)\to\Ve).
\end{align}
To understand the physical significance of the above amazing result, let us
consider a physical problem: given a system with a symmetry whose ground
state is a product state with the symmetry, if we only measure the system via
probes that do not break the symmetry, can we determine the symmetry group of
the system? Here symmetric probes correspond to operators $O$ that commute with
all group actions, $gOg^{-1}=O,\forall g$. Generic group actions are not
symmetric probes, unless they are in the center of the group. On the other hand, the fusion and braiding of the point-like
excitations above the ground state correspond to symmetric operation.  The
representation category $\Rep(G)$ contains only those symmetric probes.
Tannaka duality tells us that we can indeed determine the symmetry group via
only symmetric probes. Although the fiber functor seems to break the symmetry
if we realize it physically, mathematically it is proven that such fiber
functor always exists and is unique up to isomorphisms. Therefore, from the
data of symmetric probes (fusion and braiding) in $\Rep(G)$, we can obtain
(formally calculate) the
group $G$ up to isomorphisms, without really breaking the symmetry of the
system.

Now let us try to break the abstract theorem into more explicit terms.
Firstly, the fiber functor means nothing but realizing the abstract fusion and
braiding in $\Rep(G)$ category with the tensor product and (trivial) braiding
of concrete Hilbert spaces in a quantum system. It is helpful to consider how
we build $\Rep(G)$ in $\Ve$: a group representation is a vector space $V$
equipped with a group action $\rho_V:G\to\mathrm{GL}(V)$.  Moreover, there is a
monoidal structure for the representations, which is taking the tensor product
of the vector spaces $V\otimes_\C W$ and the new group action is
$\rho_{V\otimes_\C W}(g)=\rho_V(g)\otimes_\C\rho_W(g)$ (which is called the
fusion of group representations).

The Tannaka duality goes exactly the other direction.  Assuming that we know a
representation category $\Rep(G)$, which contains only information on symmetric
operations such as how the representations fuse with each other, can we obtain
the group actions and also the group?
The answer of the theorem then goes:
\begin{enumerate}
  \item If we have a collection of invertible linear maps $\alpha_X$ for each
    irreducible representation $X$, acting on the vector space
    $F(X)$
    assigned to $X$ by a fiber functor $F$, such that
\item They are compatible with the fusion, in the sense that $\alpha_{X\otimes
  Y}=\alpha_X\otimes_\C \alpha_Y$,
  \begin{align}
    \xymatrix{F(X\otimes Y)\ar@{=}[r]^-\sim\ar[d]^{\alpha_{X\otimes Y}} &F(X)\otimes_\C F(Y)
    \ar[d]^{\alpha_X\otimes_\C\alpha_Y}\\
    F(X\otimes Y)\ar@{=}[r]^-\sim &F(X)\otimes_\C F(Y) }
  \end{align}
It is possible that $X\otimes Y$ is a reducible representation. We extend the
 linear maps $\alpha_W$ to $W$ being reducible representations by direct sums, \ie if $W$
 is the direct sum of
irreducible representations $W_i$, $W=\bigoplus_i W_i$, $\alpha_{W}$
is given by the corresponding direct sum $\alpha_W=\bigoplus_i \alpha_{W_i} $.
\end{enumerate}
This collection of invertible linear maps $\alpha_X$, must correspond to the action
of some group element $g\in G$, $\alpha_X=\rho_{F(X)}(g)$.

Moreover, take all collections
of such invertible linear maps, they form a group under composition, namely the automorphism
group of the fiber functor, ${\Aut(F:\Rep(G)\to\Ve)}$. It is isomorphic to $G$.
In other words, if at the beginning we are given an abstract bosonic SFC $\sE$, with a
fiber functor $F:\sE\to\Ve$, we can use the above reconstruction to extract the
group underlying $\sE$, via ${\sE\cong \Rep(\Aut(F:\sE\to\Ve))}$.


%
%

\subsection{Fusion of boundary strings recover the group }

Let us focus on the loop excitations on the string-only boundary.
A loop excitation shrunk to a point
may become a direct sum of point-like excitations
(see \eqn{shrink})
\begin{align}
 s = n \one\oplus \cdots
\end{align}
where $\one$ is the trivial point-like excitation and $\cdots$ represent other
possible nontrivial point-like excitations.  When $n=0$, the string is not
pure. Another possibility is that $n>1$. In this case the string is unstable;
it has accidental degeneracy which can be lifted by perturbations. 
So the pure simple strings have $n=1$.

Since there is only trivial particle on the boundary, when we shrink a loop on
the boundary, it must become a multiple of the trivial particle, $n\one$. Thus,
it suffices to consider only the \emph{simple} loops ($n=1$) on the boundary,
which shrink to the trivial particle $\one$. In other words, simple loops on
the boundary shrinks to \emph{nothing}; this is an essential property in the
following discussions. We note that such simple loops have a quantum dimension
$d=1$, and their fusion is group-like.  For the moment, we denote the group
formed by the simple loops on the boundary under fusion (see
Fig.~\ref{strfuse}), by $H$. 

\begin{figure}[tb]
  \centering
  \includegraphics[scale=0.5]{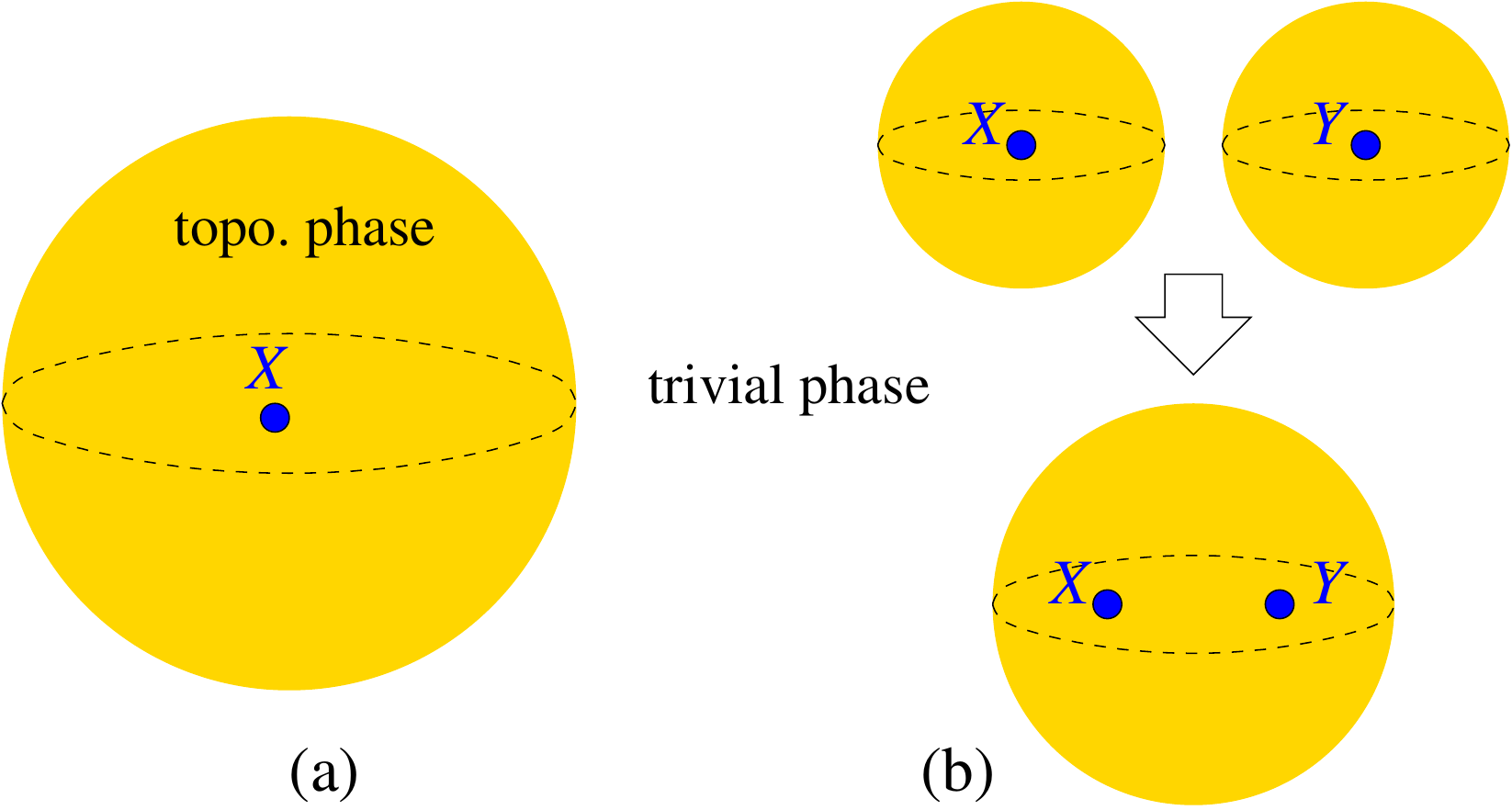}
  \caption{(Color online) (a) The fusion space $F(X)$ for a 3-disk $D^3$ containing only one
particle $X$. (b) Merging two 3-disks to one 3-disk induces an isomorphism
$F(X)\otimes_\C F(Y)\cong F(X\otimes Y)$.}
  \label{fig:fiber}
\end{figure}

To apply the Tannaka duality, we need a physical realization of the fiber
functor. Consider a simple topology for a string-only boundary: put the 3+1D
topological order $\sC^4$ in a 3-disk $D^3$, the boundary on $\partial D^3=S^2$, and
outside is the trivial phase $\sD^4$. When there is only a particle $X$ in the
3-disk,
with no string and no other particles, we associate the corresponding fusion
space (the physical states with such a configuration) to the particle $X$, and
denote this fusion space by $F(X)$ (see Fig.~\ref{fig:fiber}). Viewed from very far away, a 3-disk containing a particle $X$ is like a ``local
excitation'' in the trivial phase, thus $F(X)$ mimics a
local Hilbert space. When there are two 3-disks, each containing
only one particle, $X$ and $Y$ respectively, the fusion space is $F(X)\otimes_\C F(Y)$.
Moreover, as adiabatically deforming the system will not change the fusion
space, we can ``merge'' the two 3-disks to obtain one 3-disk containing one
particle $X\otimes Y$. Therefore $F(X)\otimes_\C F(Y)\cong F(X\otimes Y)$.
Similarly, $F$ also preserves the braiding of particles. In other words, the
assignment $X\to F(X)$ gives rise to a fiber functor. By Tannaka duality, we
can, at least formally, reconstruct a group $G=\Aut(F)$, such
that the particles in the bulk $\sC^4$ are identified with $\Rep(G)$. 
Our goal is to show that the fusion
group $H$ of the simple loops on the boundary, is the same as $G$.

\begin{figure}[tb]
  \centering
  \includegraphics[scale=0.5]{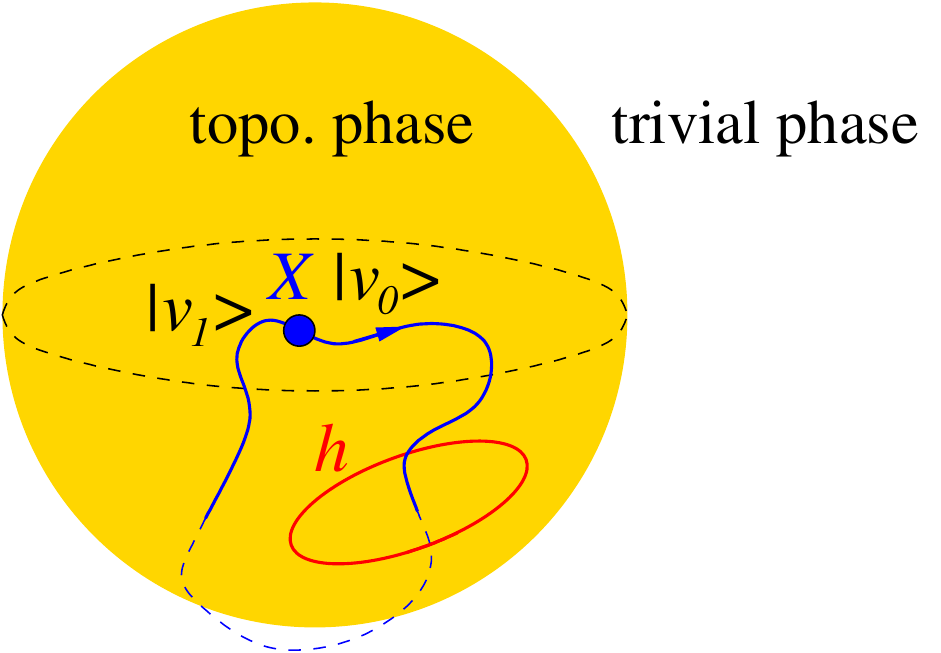}
  \caption{(Color online) Moving a particle (blue) around a loop excitation
(red) on the boundary. The solid line is a half-braiding path. The dashed line
is the complementing path in the trivial phase.}
  \label{halfbraid}
\end{figure}

To do this we consider the process of adiabatically moving a particle $X$
around a simple loop $h\in H$ on the boundary, as shown in Fig.~\ref{halfbraid}.
As the simple loop shrinks to nothing, inserting simple loops will not change the fusion
space. But an initial state $|v_0\>\in F(X)$, after such an
adiabatically moving process, can evolve into a different end state $|v_1\>\in
F(X)$. Thus,
braiding $X$ around $h$ induces an invertible (since we can always move $X$
backwards) linear map on the fusion space $F(X)$, $\alpha_{X,h}:|v_0\> \mapsto
|v_1\>$.

Next, consider that we have two particles $X,Y$ in the bulk. If we braid them
together (fusing them to one particle $X\otimes Y$) around the simple loop $h$, we obtain the linear map $\alpha_{X\otimes Y,h}$.
If the fusion of the bulk particles is given by $X\otimes Y =
\bigoplus_i W_i$, we can split $X\otimes Y$ to the irreducible representations $W_i$, and
braid
$W_i$ with $h$; in other words,
$ \alpha_{X\otimes Y,h} =
\bigoplus_i \alpha_{W_i,h}.$

But
it is also equivalent if we move $X,Y$ one after the other. More precisely, we can first separate $Y$ into another
3-disk, braid $X$ with $h$, and then merge $Y$ back to the original
3-disk.  Thus moving $X$ alone corresponds to the linear map $\alpha_{X,h}\otimes_\C
\id_{F(Y)}$. Similarly, moving $Y$ alone corresponds to
$\id_{F(X)}\otimes_\C\alpha_{Y,h}$ and in total
we have the linear map ${\alpha_{X,h}\otimes_\C\alpha_{Y,h}}$. Therefore, $\alpha_{X\otimes Y,h}=\alpha_{X,h}\otimes_\C
\alpha_{Y,h}$, or using only irreducible representations,
\begin{align}
  \alpha_{X,h}\otimes_\C \alpha_{Y,h} =
  \bigoplus_i \alpha_{W_i,h}.
\end{align}
These linear maps are compatible with the fusion of bulk
particles.



Moreover, the simple loop $h$ provides such an invertible linear map
$\alpha_{X,h}$ for \emph{each} particle $X\in\Rep(G)$ in $\sC^4$, thus by
Tannaka duality, these linear maps must correspond to the action of certain
group element $\varphi(h)\in G$, $\alpha_{X,h}=\rho_{F(X)}(\varphi(h))$. In
other words, we obtain a map $\varphi$ from the simple loops $H$ to $G$,
$\varphi:H\to G$. It is compatible with the fusion of simple loops, because the
path of braiding around two concentric simple
loops, $g_1,g_2$ (as in Fig.~\ref{strfuse}), separately, can be continuously deform to the
braiding path around the two loops together, or around their fusion
$g_1\otimes g_2=g_1 g_2$.
This implies that $\varphi(g_1)\varphi(g_2)=\varphi(g_1 g_2)$, namely,
$\varphi$ is a group homomorphism.


What we really want is that $\varphi$ is an isomorphism and $H=G$.  This is a
consequence of the remote detectability condition. Before proving it, we
explain in detail the principle of remote detectability near the string-only
boundary. The general idea is the same, that everything must be detectable
remotely. Near a string-only boundary, the only way to perform remote detection
is the \emph{half-braiding} between bulk particles and boundary strings
\cite{L1309,KW1458}. Therefore, \myfrm{(1) there is no nontrivial boundary
string that has trivial half-braiding with all the bulk particles (boundary
strings are detectable by bulk particles).\medskip\\(2) there is no nontrivial
bulk particle that has trivial half-braiding with all the boundary string.} One
may have doubts in (2): even if bulk particles can not be detected by boundary
strings,  we may still have bulk strings to detect them. The reason for (2) is
that we believe generalized boundary-bulk duality, that the bulk strings can
always be viewed as certain ``lift'' of boundary strings to the
bulk.\cite{KW1458,KW170200673} If a bulk particle has trivial half-braiding
with all boundary strings, it also has trivial braiding with all the ``lift''
of boundary strings, \ie all the bulk strings, which conflicts with the remote
detectability condition in the bulk.

A typical half-braiding path is shown in Fig.~\ref{halfbraid}. It is important
to note that the (non-Abelian) geometric phase depends on the half-braiding
path; however, we can extract a universal path-independent half-braiding
invariant, by complementing the half-braiding path into a full loop with
another half loop of path in the trivial phase outside the boundary.  Different
half loop of path in the trivial phase with the same starting and end points on
the boundary always contribute the same geometric phase (because closed paths
in the trivial phase has no geometric phase).  This way we obtain the
half-braiding invariant as the expectation value of such whole loop
adiabatically moving process (half in the bulk, half in the trivial
phase).\footnote{It is similar in 2+1D topological orders. The topological
$S$-matrix is the invariant of braiding, but it is the expectation value of
\emph{double} braidings, \ie moving one particle a whole loop around another.}
Trivial half-braiding means that such half-braiding invariant is trivial.
Immediately we see that the linear maps
$\alpha_{X,h}$ are directly related to the
half-braidings, in the sense that
$\<\alpha_{X,h}\>$ gives the above half-braiding invariant.
If $\alpha_{X,h}$ is the identity map, it implies trivial half-braiding between
$X$ and $h$.

Now, we are ready to show that $\varphi:H\to G$ is an isomorphism:
\begin{enumerate}
  \item $\varphi$ is injective. Consider $\ker\varphi$, namely the simple loops
    that induce just identity linear maps on all bulk particles. In
    other words, $\ker\varphi$ consists of simple loops that have trivial
    half-braiding with all bulk particles. By the remote detectability condition
    (1), $\ker\varphi$ must be trivial, which means $\varphi$ is injective.

  \item $\varphi$ is surjective.
    We already showed that $\varphi:H\to G$ is injective, so we can view $H$ as
    a subgroup of $G$.

    Now consider a special particle in the bulk, which
    carries the representation $\mathrm{Fun}(G/H)$,
    linear functions on the right cosets $G/H$. More precisely,
    $\mathrm{Fun}(G/H)$ consists of all linear functions on $G$, $f:G\to \C$,
    such that $f(h x)=f(x),$ $\forall h\in H,x\in G$ (takes the same value on
    a coset).
    The group action is the usual one on functions,
    ${\rho_{\mathrm{Fun}(G/H)}(g):f(x)\mapsto f(g^{-1} x)}$.

    The linear maps $\alpha_{X,h}$ induced by the simple loops
    are all actions of group elements in $H$,
    and they are all identity maps on the
    special particle $\mathrm{Fun}(G/H)$. In other words, the bulk
    particle $\mathrm{Fun}(G/H)$ has trivial half-braiding with all the
    boundary strings. By the remote detectability condition (2), it
    must be the trivial particle carrying the trivial representation.
    In other words, we have $G=H$.
\end{enumerate}

To conclude, the simple loop excitations on the string-only boundary, forms a
group under fusion. It is exactly the same group whose representations are
carried by the point-like excitations in the bulk.

If we insert a bulk loop excitation in the 3-disk and perform a similar
braiding process, it also induces a linear map on the underlining fusion space.
One may wonder if this also associates group elements to the bulk strings. This
is in general not true.
Unlike the boundary simple loops, inserting a bulk string, even if it is pure
and simple, will enlarge the fusion space of only particles, as long as the
quantum dimension of such string is greater than 1. As a result, only those
bulk strings with quantum dimension $d=1$
can be associated with group elements. In section \ref{untw} we have shown that
all the bulk strings can be associated with conjugacy classes of the group $G$
(even if topological order is not a $G$-gauge theory). Some further discussions
can be found in Appendix \ref{appendix:str}.

\section{The classification of unitary pointed fusion 2-categories}

\subsection{A mathematical formulation}

First, let us consider the so-called unitary pointed fusion \mbox{(1-)categories}.  A pointed
fusion category consists of a finite number of simple objects. A simple object $x$ is an object such that Hom$(x,x)=\C$. For each simple object $x$, there is a simple object $y$ such that $x
\otimes y = 1$, where $1$ is the tensor unit, \ie $ 1\otimes x = x = x\otimes
1$.  In other words, the set of simple objects form a finite group $G$. We will also
denote the simple object by $g_1$, $g_2$, $g_3$, etc.

In this case, the only not-yet-fixed structure is the associator isomorphism: 
\begin{align}
(g_1 \otimes g_2) \otimes g_3 \to g_1 \otimes (g_2 \otimes g_3) .
\end{align}
Note that both domain and target are the same simple object. But the associator
isomorphisms can be nontrivial. By  the simpleness, the isomorphism is just a
non-zero complex number $\om_3(g_1, g_2, g_3)$. If the theory is unitary, one
needs to further require that this number is a phase (in $U(1)$).  Then the
pentagon condition implies that $\om_3(g_1, g_2, g_3)$ is a 3-cocycle. Different
cocycles give equivalent fusion categories if they differ by a 3-coboundary.
Moreover, we may permute the simple objects by group automorphisms, thus two different
cocycles also give equivalent fusion categories if they got mapped to each
other by group automorphisms.
In conclusion, unitary pointed fusion categories,
are classified by $(G,\om_3)$ where $\om_3\in\cH^3[G; U(1)]$ up to group
automorphisms.

Now let us consider a pointed unitary fusion 2-category \cite{Gurski07}. We
will not define it in full detail here, but only describe some physically
relevant ingredients of it.  It has only finite number of simple objects. A
generic object is a direct sum of simple objects. For two simple objects $x,y$,
we have $\Hom(x,x)=\Ve$ and $\Hom(x,y)=0$ for $x\neq y$, where $0$ is the
category consisting of only the $0$-vector space.  The tensor unit $1$ is
simple. For each simple object $x$, there is a simple object $y$ such that $x
\otimes y =1$, where $1$ is the tensor unit. So, again the set of simple
objects is a finite group $G$. We will denote simple objects by group elements
$g_1, g_2, g_3$, etc.

For a simple object $g$, the identity 1-morphism $\text{id}_g$ is $\C$ (the
only invertible object in $\Ve$). The 2-morphisms form Hom$(\text{id}_g,
\text{id}_g) = \Hom_\Ve(\C,\C)=\C$, and there are unit 1-isomorphisms and
associator 1-isomorphisms: (1) The unit 1-isomorphisms: $1\otimes g = g \to g$
is just the identity morphism $\id_g=\C$.  (2) The associator 1-isomorphism:
\begin{align} (g_1\otimes g_2) \otimes g_3 \to g_1 \otimes (g_2 \otimes g_3)
\end{align} is also still the identity 1-morphism $\id_{g_1g_2g_3}=\C$.  
There are two ways to go from $((g_1 g_2)g_3)g_4$ to $g_1(g_2(g_3g_4))$ via
identity 1-morphisms.  Therefore, two paths both give the identity map
$\text{id}_{g_1g_2g_3g_4}=\C$.  So the commutative of the pentagon diagram is
clear.  But we can introduce for each pentagon a 2-isomorphism: $\C\to \C$,
which is a phase, denoted by $\om_4(g_1,g_2,g_3,g_4)$.  These 2-isomorphisms
need satisfy a higher coherence relation.  Then this coherence relation implies
that $\om_4(g_1,g_2,g_3,g_4)$ is a 4-cocycle. Again, 4-cocycles differing only
by a coboundary give equivalent pointed fusion 2-categories.  One can do the
same for the triangle relation.  Namely, one can introduce a 2-isomorphism for
each triangle. We believe that these 2-isomorphisms should give the same
unitary fusion 2-category.  

The same structure is discussed in \Ref{EN170202148}, under a
different name, $G$-graded 2-vector spaces $\mathbf{2Vec}^{\om_4}_G$, where
they also believe that $(G,\om_4)$ is enough to determine
a unitary pointed fusion 2-category.

Note that the equivalence between unitary pointed fusion 2-categories must
preserve the tensor product of simple objects, thus must correspond to some group automorphism ${\phi:G\cong G}$. Such automorphism also
acts on the cocycles (necessarily change the cocycle if it is an outer
automorphism, namely, not of the form $x\mapsto gxg^{-1}$ for some $g\in G$).
Under such automorphism $\phi$, $(G,\om_4)$ and $(G,\phi(\om_4))$, where
$\om_4$ is a $4$-cocycle, correspond to the same pointed unitary fusion
$2$-category.  Therefore, we believe that pointed unitary fusion 2-categories
one-to-one correspond to the pairs $(G,\om_4)$ where $\om_4\in\cH^4[G; U(1)]$, up to group
automorphisms.
 

\subsection{A physical argument}

\begin{figure}[tb] 
\centering \includegraphics[scale=0.4]{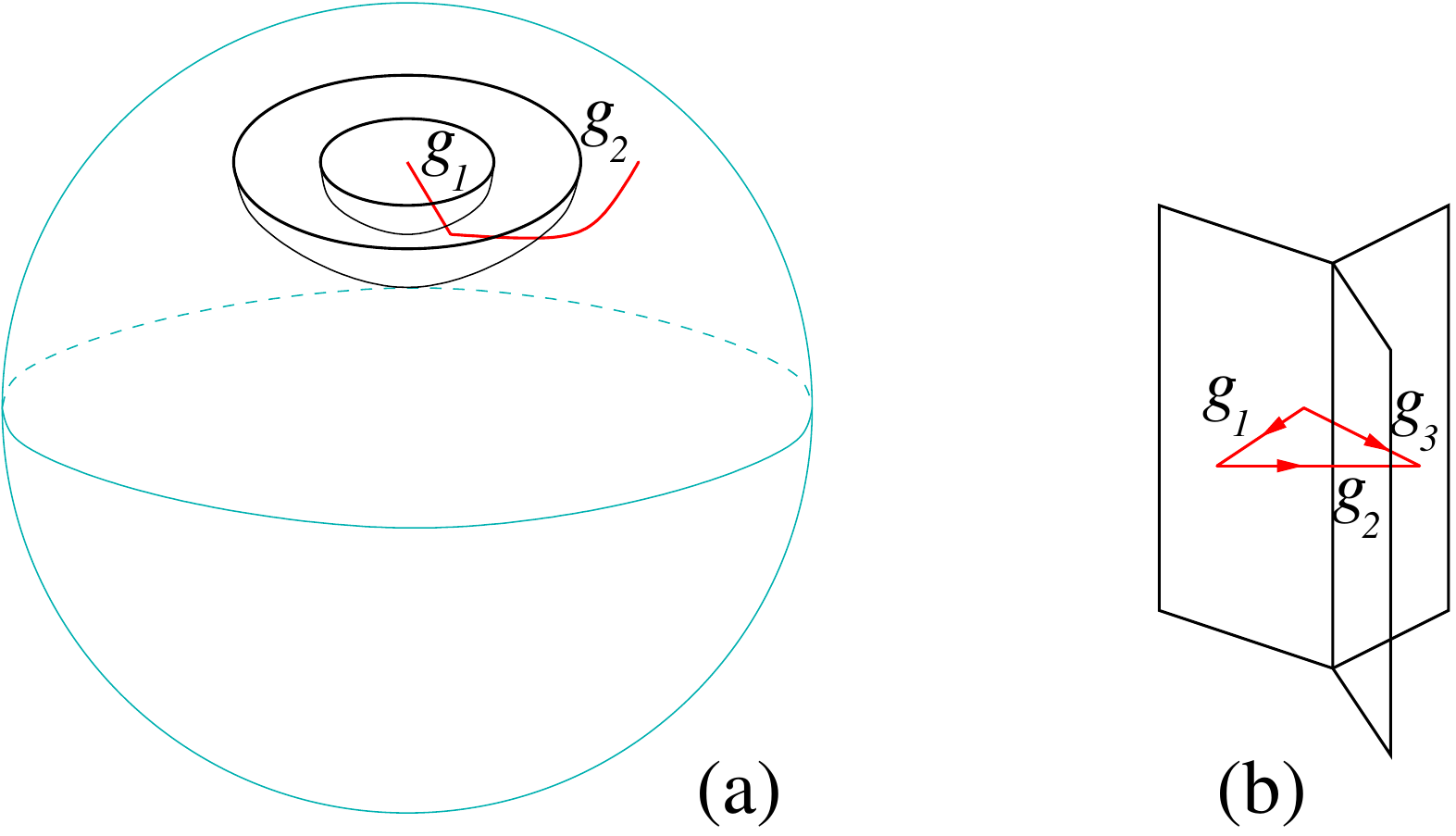} 
\caption{
(Color online)
(a) Two boundary strings $s^\text{bdry}_{g_1}$ and $s^\text{bdry}_{g_2}$ on the
surface $S^2$ of $D^3$.  The process of fusing the two strings to a no string
state is described by a membrane-net in $D^3$ (which is formed by two
hemispheres in this case).  The dual of the  membrane-net is a string-net
(which is formed by two strings in this case).  (b) The intersection of three
membranes is dual to a triangle of strings.  The fusion rule of the boundary
strings corresponds to the flat connection condition \eqn{flat} on the
string-net.
}
\label{strfuseS2} 
\end{figure}

In the following, we will try to understand the above mathematical result from
a physical point of view.  Let the 3-dimensional space to be a 3-disk $D^3$.
Consider the boundary strings $s^\text{bdry}_{g_1}$ and $s^\text{bdry}_{g_2}$
on the surface of the 3-disk $S^2=\prt D^3$ (see Fig. \ref{strfuseS2}a).  The
process for the boundary strings to fuse to a non-string state can be
represented by a membrane-net in $D^3$ (see also \Ref{EN170202148}).  The same
boundary strings can fuse to a non-string state through a different process
which is represented by another membrane-net in $D^3$.  To compare the two
processes, we can glue the boundary of the above two membrane together along
the $S^2$, to form a membrane-net in $S^3$.  Such a membrane-net in $S^3$
describe the process of creating boundary strings from a no-string state, and
then fuse those boundary strings to no-string state.  Such a process induce a
$U(1)$ geometric phase $\ee^{\ii \th}$, since the fusion space of the boundary
strings is always 1-dimensional.  So we assign such a $U(1)$ geometric phase
$\ee^{\ii \th}$ to the membrane-net on $S^3$.

But such a $U(1)$ geometric phase may not have a local expression.  Let us
assume that the membrane-net on $S^3$ is formed by the 2-simplices of a
triangulation of $S^3$.  The vertices of the triangulation are labeled by
$I,J,K,\cdots$.  ``Having no local expression'' means that we cannot assign a
phase factor $\om_3(IJKL)$ to each 3-simplex $\<IJKL\>$ of the triangulation to
express the total $U(1)$ geometric phase $\ee^{\ii \th}$ as a product of those
local phases:
\begin{align}
 \ee^{\ii \th} \neq \prod_{\<IJKL\>} \om_3(IJKL).
\end{align}


\begin{figure}[tb] 
\centering \includegraphics[scale=0.8]{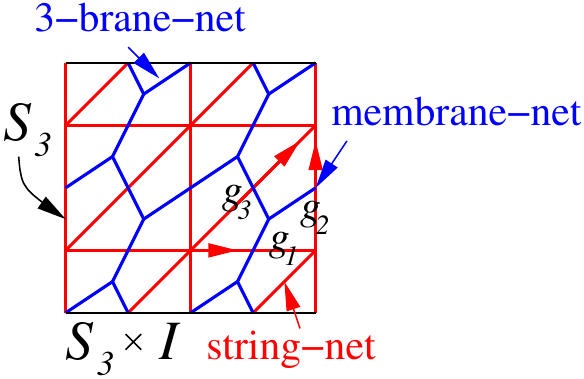} 
\caption{
(Color online)
Two processes are described by membrane-net on the two boundaries of $S^3\times
I$.  The change between the two processes is described by a 3-brane-net (the
blue lines) on $S^3\times I$ (with 2 of the 3 dimensions suppressed).  The red
lines form the string-net which is dual to the 3-brane-net.
}
\label{brane} 
\end{figure}

We see that the process of creating some boundary strings from nothing and then
fusing them to nothing can be represented by a membrane-net on space $S^3$.
Such a process correspond to a phase factor $\ee^{\ii \th}$.  Two different
processes of creating some boundary strings from nothing and then fusing them
to nothing give rise to two phase factors $\ee^{\ii \th}$ and $\ee^{\ii \th'}$.
The two processes can be compared by a ``time''-evolution from the membrane-net on
$S^3$ that correspond to the first process, to the membrane-net on $S^3$ that
correspond to the second process.  In other words, the comparison of the two
processes is represented by a 3-brane-net on $S^3\times I$, where $S^3$ is the
space and the segment $I$ represents the ``time'' direction (see Fig. \ref{brane}).
The first process corresponds to the membrane-net on one boundary of $S^3\times
I$ which is one boundary of the 3-brane-net on $S^3\times I$.  The second
process corresponds to the membrane-net on the other boundary of $S^3\times I$
which is  the other boundary of the 3-brane-net  on $S^3\times I$.  

In 4-dimensions, a 3-brane-net is dual to a string-net where each 3-brane in
the 3-brane-net intersects with a string in the string-net (see Fig.
\ref{brane} and \ref{strfuseS2}). So the strings in the string-net is also
labeled by $g_i$.  In the 3-brane-net, only the 3-branes that satisfy the
fusion rule \eqn{sssbdry} can intersect along a line (see Fig. \ref{brane} and
\ref{strfuseS2}b).  This means that the labels of the strings in the string-net
satisfies
\begin{align}
\label{flat}
 g_1g_2=g_3.
\end{align}
[Note the same string with opposite orientations is labeled by $g$ and $g^{-1}$
respectively. The orientation of strings in the string-net is chosen to from a
branching structure (see Appendix \ref{branch}) of the string-net .] The above
happen to be the flat connection condition if we view $g_i$ on a string as the
gauge connect between the two vertices connected by the string.  So the
evolution from one process to the other can be represented by a string-net on
$S^3\times I$.

The two different processes may differ by a phase factor $\ee^{\ii (\th' -
\th)}$.  So we can assign the string-net on $S^3\times I$ by such a phase
factor, to represent the difference of the two processes.  Let us assume the
string-net on  $S^3\times I$ is formed by the edges of a triangulation of
$S^3\times I$. Then starting from one boundary of $S^3\times I$, we can build
the whole triangulation of $S^3\times I$ by adding one pentachoron (\ie one
4-simplex) at a time.  We note that adding a pentachoron corresponds to change
one process to its neighboring process.  The difference of the two neighboring
processes is described by the added pentachoron with edges labeled by $g_{IJ}$,
$I,J=0,1,2,3,4$ (where $I=0,1,2,3,4$ label the five vertices of the
pentachoron).  We may assign the phase difference of the two neighboring
processes to the added pentachoron.  So each pentachoron is assigned to a phase
factor $\om_4(\{g_{IJ}\})$.  Due to the flat connection condition \eqn{flat},
$g_{IJ}$'s for the pentachoron are not independent. So $\om_4(\{g_{IJ}\})$ can be
rewritten as $\om_4(g_{01}, g_{12}, g_{23},g_{34})$.  Such a 4-variable
function on $G$ can be viewed as a group 4-cochain.

So the total phase difference of the two processes can be written as
\begin{align}
 \ee^{\ii (\th' - \th)}=\hskip -1.5em \prod_{\<IJKLM\> \in S^3\times I} \hskip -1.5em \om_4^{s_{IJKLM}}(g_{IJ}, g_{JK}, g_{KL},g_{LM})
\end{align}
where $\prod_{\<IJKLM\> \in S^3\times I}$  multiply over all the pentachorons
$\<IJKLM\>$ in the triangulation of $S^3\times I$, and $s_{IJKLM}=\pm 1$
describes the two different orientations of the pentachorons $\<IJKLM\>$ which
arises from  the branching structure (see Appendix \ref{branch}).

If we choose the two processes described by the boundary of $S^3\times I$ to be
the do nothing process that the leave the no-string state unchanged, then the
string-net on $S^3\times I$ can be viewed as a string-net on $S^4$.  The total
phase difference of the two do-nothing processes (which is actually the same
process) should be zero:
\begin{align}
 1 = \prod_{\<IJKLM\> \in S^4}  \om_4^{s_{IJKLM}}(g_{IJ}, g_{JK}, g_{KL},g_{LM}),
\end{align}
and the above should hold for any triangulation of $S^4$ and any assignment of
the label $g_{IJ}$ on the edges (as long as the flat connection condition
\eqn{flat} is satisfied).  It implies that $\om_4(g_{01}, g_{12},
g_{23},g_{34})$  is a group 4-cocycle in $\cH^4(G;U(1))$.  
This is a physical way
to explain why \myfrm{Unitary pointed fusion 2-categories are classified by a
pair $(G,\om_4)$ up to group automorphisms, where $G$ is a finite group and
$\om_4$ its group 4-cohomology class: $\om_4 \in \cH^4(G;U(1))$.}

\section{From boundary to bulk}

We have shown that all 3+1D AB bosonic topological orders 
can have a boundary described by pointed unitary fusion
2-category $\cM^3$ whose fusion is given by the group $G$.  It is believed the
boundary anomalous topological order completely determine the bulk topological
order \cite{KW1458,KW170200673}.  
More precisely, the bulk topological order should be given by the center $Z(\cM^3)$,
which can be explicitly defined by the 2-category $\mathcal{F}un_{\cM^3|\cM^3}(\cM^3,\cM^3)$ 
of $\cM^3$-$\cM^3$-bimodule 2-functors. 

But their relation can be many-to-one:
several different anomalous boundary topological orders may correspond to the
same bulk topological order; in other words, the same bulk topological order
can have several different gapped boundaries. 
Since 3+1D topological orders always have gapped
boundaries, all 3+1D topological orders are determined by some anomalous
2+1D boundary topological orders. Mathematically, we say that there is
surjective map from the set of anomalous 2+1D boundary topological orders
to the set of 3+1D topological orders
\begin{align}
&\ \ \ \
\text{2+1D boundary anomalous topological orders}
\nonumber \\
& 
\twoheadrightarrow 
\text{3+1D topological orders} .
\end{align}
Furthermore, since all 3+1D AB topological orders have a string-only boundary
described by unitary pointed fusion 2-categories, we further have
\begin{align}
 &\ \ \ \
\text{Unitary pointed fusion 2-categories}
\nonumber \\
& 
\twoheadrightarrow 
\text{3+1D AB topological orders} .
\end{align}

In this paper the string-only boundary is obtained by condensing all the
point-like 
particles that form $\Rep(G)$. A natural question is whether such
condensation process, and also the resulting boundary, are unique or not.

Firstly, we believe that the condensation of particles in 3+1D follows the same
rule as that for condensation of anyons in 2+1D (at least if we restrict the
3+1D condensation to a 2+1D sub-manifold, see Fig. \ref{3D2Dsub}). Anyon condensation in 2+1D has been
thoroughly studied. It is fully controlled by the so-called
\emph{condensable algebra}\cite{Kon14} in the category of anyons. In other
words, the condensable algebra completely determines the condensed phase and
the domain wall/boundary between the old phase and the condensed phase.

Thus, we should focus on the condensable algebras in $\Rep(G)$. They are
already classified in Theorem 2.2 in \Ref{KO01} (see also Theorem 3.7 in
\Ref{LKW1602.05936}). There is a unique condensable algebra that condenses all the particles in $\Rep(G)$. It is given by $\mathrm{Fun}(G)$, the algebra of all functions on $G$.  Therefore, there is only one way to condense all particles
$\Rep(G)$. We obtain a unique condensed phase, which is trivial. As result,
there is a unique, also canonical, string-only boundary.

This way, we got an even stronger result.  Each 3+1D AB topological order only
have a unique boundary that corresponds to the condensation of all point-like
excitations.  In other words,  each 3+1D topological order corresponds to a
unique unitary pointed fusion 2-category.  \myfrm{Unitary pointed fusion
2-categories classify all 3+1D AB topological orders in a one-to-one fashion.}
Such a result is similar to a result in one lower dimension: \myfrm{Unitary
fusion categories describes all 2+1D topological orders with gappable boundary
(but in a many-to-one way) \cite{LW0404617,LW1384}.}

Let us briefly explain why the approach used in this paper for 3+1D topological
orders does not apply in 2+1D, which gives a flavour why we
can obtain a stronger result in 3+1D. In 3+1D, all point-like excitations have
trivial statistics; if they are all bosons, it is a natural and canonical
choice to condense all of them and we obtain a unique string-only 2+1D gapped
boundary. In 2+1D, there are only point-like excitations with
non-trivial statistics between them. One can similar choose a subset of
quasiparticles to condense; if the subset is big enough one can also obtain a
gapped 1+1D boundary. However, in general there are several such subsets to
condense, among which none is special. As a result, there is no canonical
gapped 1+1D boundary. This essential difference makes the
classification of topological orders in
3+1D simpler than those in 2+1D.

\section{Realization by Dijkgraaf-Witten models}

Combining the results from the last a few sections, we obtain that \myfrm{3+1D
AB topological orders are classified by a finite group $G$ and its
group 4-cocycle $\om_4 \in \cH^4(G;U(1))$, up to group automorphisms.} A finite
group $G$ and its group 4-cocycle happen to be the data needed to construct the
Dijkgraaf-Witten model.  In fact all the 3+1D AB topological orders can be
realized by Dijkgraaf-Witten models.

We note that 3+1D Dijkgraaf-Witten models \cite{DW9093} are defined on a
4-dimensional simplicial complex with branching structure (see Appendix
\ref{branch}).  Let us use $I,J,\cdots$ to label the vertices of the complex.
The degrees of freedoms live on the links of the complex, which are labeled by
$g_{IJ} \in G$ where $G$ is a finite group.  $g_{IJ}$'s satisfies a
flat-connection condition
\begin{align}
 g_{IJ}g_{JK}=g_{IK},
\end{align}
for any triangles $\<IJK\>$. The Dijkgraaf-Witten models are defined via a path
integral
\begin{align}
 Z =\sum_{\{g_{IJ}\}} \prod_{\<IJKLM\>}
\om_4^{s_{IJKLM}}(g_{IJ},g_{JK},g_{KL},g_{LM}),
\end{align}
where $\prod_{\<IJKLM\>}$ multiply over all the 4-cells $\<IJKLM\>$ whose vertices
are ordered as $I<J<K<L<M$.  Also, $s_{IJKLM} =\pm 1$ describes the orientation
of the 4-cell $\<IJKLM\>$ (see Appendix \ref{branch}), and $\om_4$ is a group
4-cocycle $\om_4 \in \cH^4[G;U(1)]$.

When the space-time has a boundary, we can obtain an exactly soluble boundary
by setting $g_{IJ}=1$ on all the links $\<IJ\>$ on the boundary.  Such an exactly
soluble boundary is actually the string-only boundary discussed in this paper.
The world-lines of topological point-like excitations are described by Wilson
lines in the bulk
\begin{align}
 \prod_{\<IJ\>} R(g_{IJ})
\end{align}
where $R$ is a representation of $G$.  But on the boundary, $g_{IJ}=1$ and
$R(g_{IJ}=1)$ is an identity matrix.  All the different topological point-like
excitations becomes the same trivial excitation on the boundary.  However,
there are non-trivial string-like excitations on the boundary. The world-sheet
of those boundary string-like excitations is given by the following: Draw a
membrane on the 3-dimension boundary of space-time.  Change $g_{IJ}$ on the
links that intersect the membrane from $g_{IJ}=1$ to $g_{IJ}=h$.  Such a change
still satisfy the flat-connection condition.  We see that different boundary
strings are labeled by the group elements and their fusion is given by the
group multiplication.  Therefore, Dijkgraaf-Witten models can realize all
unitary pointed fusion 2-category on the boundary.  Using the boundary-bulk
relation \cite{KW1458,KW170200673}, we can show that Dijkgraaf-Witten models can
realize all 3+1D AB topological orders.

\section{Relation to 3+1D bosonic SPT orders}

There are two kinds of SPT orders when the symmetry group is unitary and
finite: the ones whose boundary have a pure gauge-anomaly will be called
pure SPT orders \cite{CGL1314,CGL1204}, and the ones whose boundary have a
mixed gauge-gravity-anomaly will be called mixed SPT orders \cite{W1477}.
In 3+1D space-time, the pure SPT orders are classified by group
cohomology $\cH^4[G;U(1)]$, while all the mixed SPT orders are described by
\emph{some} elements in \cite{W1477}
\begin{align}
&\ \ \ \
\cH^1(G;\cH^3[SO(\infty);U(1)]) \oplus \cH^2(G;\cH^2[SO(\infty);U(1)])
\nonumber\\
&= \cH^1(G;\Z) \oplus \cH^2(G;\Z_2) = \cH^2(G;\Z_2).
\end{align}
For many groups, $\cH^2(G;\Z_2)\neq 0$. But a non zero $\cH^2(G;\Z_2)$ does not
implies the existence of mixed SPT, since not all the elements in
$\cH^2(G;\Z_2)$ correspond to existing SPT orders.

Since 3+1D AB topological orders can be obtained by gauging \cite{LG1209} the
symmetry of 3+1D bosonic SPT states, and since Dijkgraaf-Witten models only
correspond to gauging the pure SPT states, we see that the classification
results in this paper implies that \myfrm{In 3+1D, there is no mixed bosonic
SPT order for unitary finite symmetry group $G$.} 

In fact, using SPT invariant, we can directly show that for unitary
finite symmetry group $G$, there is no mixed SPT orders in 3+1D. (However, if
$G$ contains time reversal, there are mixed SPT orders in 3+1D
\cite{VS1306,K1459}.) To obtain SPT invariant, we gauge the symmetry and put a
flat-connection $A$ on a closed orientable space-time $M^4$. The partition
function of the system on $M^4$ with a fixed flat-connection $A$ is the so
called SPT invariant \cite{W1447,HW1339,WGW1405.7689}.  If there is a mixed 3+1D SPT
order described by $\cH^2(G;\Z_2)=\bigoplus \Z_2$, its SPT invariant will have
a form 
\begin{align}
 Z(M^4,A) = \ee^{\ii \pi \int_{M^4} \om_1(A)\smile {\rm w}_3 + \om_2(A)\smile {\rm w}_2},
\end{align}
where $\om_n(A),\ {\rm w}_n$ are topological $n$-cocycles in $H^n(M^4;\Z_2)$,
and ${\rm w}_n$ is also the $n^\text{th}$ Stiefel-Whitney class.  There are
many relations between Stiefel-Whitney classes and cocycles $\om_n(A)$.  For
example, by calculating $Sq^1(\om_1(A)\smile {\rm w}_2)$ in two different ways,
we find that on orientable $M^4$, $\om_1(A)\smile {\rm w}_3 =\om_1(A)\smile
\om_1(A)\smile {\rm w}_2$ \cite{W1477,W161201418}.  (Here $Sq^n$ is the Steenrod
operation.) Thus
\begin{align}
 Z(M^4,A) = \ee^{\ii \pi \int_{M^4} [\om_2(A) +\om_1(A)\smile \om_1(A)]\smile {\rm w}_2},
\end{align}
Similarly, $[\om_2(A) +\om_1(A)\smile \om_1(A)] \smile {\rm w}_2=Sq^2[\om_2(A)
+\om_1(A)\smile \om_1(A)]=[\om_2(A) +\om_1(A)\smile \om_1(A)]\smile[ \om_2(A)
+\om_1(A)\smile \om_1(A)]$.  Thus
\begin{align}
 Z(M^4,A) = \ee^{\ii \pi \int [\om_2(A) +\om_1(A)\smile \om_1(A)]\smile [\om_2(A) +\om_1(A)\smile \om_1(A)]}.
\end{align}
We see that the SPT order described by the above SPT invariant is actually a
pure SPT order described by $\cH^4[G;U(1)]$ \cite{K1459,W1477} and hence, there
is no mixed SPT order in 3+1D for unitary finite symmetry group.  This result
supports our classification of 3+1D AB topological orders in terms of
Dijkgraaf-Witten models.  In 4+1D, there is a mixed bosonic $Z_2$ SPT state
\cite{W1477}.  Gauging such a mixed $Z_2$ SPT state will produce a 4+1D AB
topological order that is beyond Dijkgraaf-Witten theory.

\section{Walker-Wang models and particle-only boundaries}

We like to remark that Walker-Wang models \cite{WW1132,KBS1307,WW160607144,CF14036491} is another quite
systematic way to construct 3+1D bosonic topological orders.  In fact,
Walker-Wang models realize all 3+1D bosonic topological orders who have a
particle-only boundary, which is described by a premodular tensor category.
Such particle-only boundary can exist for a 3+1D topological order, if
condensing the maximum set of strings that have trivial mutual braiding will
change the 3+1D topological order to a trivial phase.


It is known that Walker-Wang models (and the related
3+1D string-net models \cite{LW0404617}) can realize 3+1D bosonic topological
orders with emergent fermionic point-like excitations.~\cite{KBS1307}  It appears that
Walker-Wang models cannot realize all 3+1D Dijkgraaf-Witten models (\ie not all
3+1D bosonic topological orders  whose point-like excitations are all bosons).

\section{Summary}

3+1D topological orders contain both point-like and string-like excitations.
At first, it appears that 3+1D topological orders, with all the fusion and
braiding of those point-like and string-like excitations, have a very
complicated structure, which may be hard to classify.  However, in this paper,
we obtain a very simple classification of 3+1D topological orders for bosonic
systems, when all the point-like excitations are bosons: they are classified by
unitary pointed fusion 2-categories, which in turn are classified by pairs
$(G,\om_4)$ up to group automorphisms.  This gives us hope that the 3+1D
topological orders may not be that complicated.  We may get a simple
classification even for the general case when some point-like excitations are
emergent fermions. We hope that the arguments developed in this paper are
helpful for such a task, which we plan to carry out in a forthcoming work.

~

~

We like to thank Meng Cheng, Zhenghan Wang, and Edward Witten for helpful
discussions.  XGW is supported by NSF Grant No.  DMR-1506475 and NSFC 11274192.
Research at Perimeter Institute is supported by the Government of Canada
through Industry Canada and by the Province of Ontario through the Ministry of
Research.  

\appendix

\section{An example: 3+1D $G$-gauge theory}
\label{gauge}

To gain an intuitive understanding of 3+1D topological orders and to introduce
the related concepts, let us study an exactly soluble local bosonic model whose
ground state has a topological order described by a 3+1D gauge theory of a
finite group $G$.  Our lattice bosonic model is defined on a 3D spatial lattice
whose sites are labeled by $I$.  The degrees of freedom live on the links
labeled by $IJ$.  On an oriented link $IJ$, such degrees of freedom are labeled
by $g_{IJ} \in G$.  $g_{IJ}$'s on links with opposite orientations satisfy 
\begin{align}
 g_{IJ}=g_{JI}^{-1}
\end{align}
The Hamiltonian of the exactly soluble model is expressed in terms of string
operators and membrane operators.

\subsection{The string operators}

The string operators are labeled by $i$'s, the irreducible
representations $R_i(g_{IJ})$ of the gauge group $G$ (where $R_i(g_{IJ})$ is the
matrix of the irreducible representation):
\begin{align}
 B_i |\{g_{IJ}\}\> &= \Big[\Tr \prod_{IJ \in \text{string}} R_i(g_{IJ})\Big] |\{g_{IJ}\}\>\nonumber\\
&=  \Big[ \Tr R_i(\prod_{IJ \in \text{string}} g_{IJ}) \Big] |\{g_{IJ}\}\>
\end{align}
We note that
\begin{align}
 B_i B_j = 
\Tr \prod_{I \in \text{string}} R_i(g_{IJ}) \otimes_\C R_j(g_{IJ}).
\end{align}
We use $\otimes_\C$ to denote the usual tensor product of matrices or vector spaces over
 the complex numbers $\C$, while $\otimes$ to denote the fusion of excitations. 
Using 
\begin{align}
R_i(g) \otimes_\C R_j(g) =\bigoplus_k N^{ij}_k R_k(g)
\end{align}
we see that
\begin{align}
  B_i B_j = \sum_k N^{ij}_k B_k.
\end{align}
The ends of the strings are point-like topological excitations and the above
$N^{ij}_k$ are the fusion coefficients of those  topological excitations.  Let
$d_i$ be the quantum dimension of those  topological excitations which satisfy
\begin{align}
\sum_j N^{ij}_k d_j = d_id_k ,
\end{align}
and let
\begin{align}
 B = \sum_i \frac{d_i}{D^2} B_i,\ \ D^2=\sum_i d_i^2 .
\end{align}
We have
\begin{align}
 B^2 &= \sum_{i,j} \frac{d_id_j}{D^4} B_i   B_j
  = \sum_{i,j,k} \frac{d_id_j}{D^4}N^{ij}_k B_k 
  \nonumber\\
  &= \sum_{i,k} \frac{d_id_i}{D^4} d_k B_k =B.
\end{align}
Thus, $B$ is a projection operator.  In fact, it is a projection operator into
the subspace with $\prod_{{IJ} \in \text{string}} g_{IJ}=1$.


\subsection{The membrane operators} 

\begin{table*}[!ht]
\caption{Fusion rules of point-like and string-like excitations in 3+1D
$S_3$ gauge theory.  Here $p_0$ corresponds to the trivial point-like
excitations which is also the trivial string-like excitations.  $p_1$ and $p_2$
are nontrivial point-like excitations corresponding to the 1D and 2D
representation of $S_3$ (\ie the charged particles).  $s_{20}$ and $s_{30}$
correspond to pure string-like excitations labeled by conjugacy classes $\chi_2$
and $\chi_3$, and   $s_{21}$, $s_{31}$ and $s_{32}$ are charge and string
bound state, as one can see from the fusion rules. See \Ref{MW1514}.
}\label{strfusion}
\centering
\begin{tabular}{|c||c|c|c|c|c|c|c|c|}
\hline
$\otimes$&$p_0$& $p_1$ &$p_2$ 	&$s_{20}$ (pure)  &$s_{21}$ &$s_{30}$ (pure) & $s_{31}$ & $s_{32}$  \\
\hline
\hline
$ p_0 $&$p_0$&$p_1$  &$p_2$						 &$s_{20}$ 										   &$s_{21}$   									 &$s_{30}$					   &$s_{31}$	           & $s_{32}$ \\
$p_1$	 &$p_1$	 &$p_0$&$p_2$						 &$s_{21}$	      							   	   &$s_{20}$											 &$s_{30}$                      &$s_{31}$ &$s_{32}$\\
$p_2$ 	 &$p_2$	 &$p_2$  &$p_0\oplus p_1\oplus p_2$&$s_{20}\oplus s_{21}$				 				   &$s_{20}\oplus s_{21}$								 &$s_{31}\oplus s_{32}$          &$s_{30}\oplus s_{32}$&$s_{30}\oplus s_{31}$\\
$s_{20}$		 &$s_{20}$    &$s_{21}$  &$s_{20}\oplus s_{21}$				 &$p_0\oplus p_2\oplus s_{30}\oplus s_{31}\oplus s_{32}$&$p_1 \oplus p_2\oplus s_{30}\oplus s_{31}\oplus s_{32}$ &$s_{20}\oplus s_{21}$ 		   &$s_{20}\oplus s_{21}$ &$s_{20}\oplus s_{21}$\\
$s_{21}$ 	 &$s_{21}$	 &$s_{20}$ 	 &$s_{20}\oplus s_{21}$				 &$p_1 \oplus p_2\oplus s_{30}\oplus s_{31}\oplus s_{32}$ &$p_0\oplus p_2\oplus s_{30}\oplus s_{31}\oplus s_{32}$&$s_{20}\oplus s_{21}$ 		   &$s_{20}\oplus s_{21}$ &$s_{20}\oplus s_{21}$\\
$s_{30}$ 	 &$s_{30}$ 	 &$s_{30}$    &$s_{31}\oplus s_{32}$			 &$s_{20}\oplus s_{21}$						 		   &$s_{20}\oplus s_{21}$				 				 &$p_0\oplus p_1\oplus s_{30}$&$s_{32}\oplus p_2$ &$s_{31}\oplus p_2$\\
$s_{31}$ 	 &$s_{31}$  &$s_{31}$  &$s_{30}\oplus s_{32}$				 &$s_{20}\oplus s_{21}$	 						       &$s_{20}\oplus s_{21}$								 &$s_{32}\oplus p_2$          &$p_0\oplus p_1\oplus s_{31}$ &$s_{30}\oplus p_2$\\
$s_{32}$ 	 &$s_{32}$	 &$s_{32}$  &$s_{30}\oplus s_{31}$				 &$s_{20}\oplus s_{21}$								   &$s_{20}\oplus s_{21}$								 &$s_{31}\oplus p_2$          &$s_{30}\oplus p_2$ &$p_0\oplus p_1\oplus s_{32}$ \\
\hline
\end{tabular}
\end{table*}

A membrane is formed by the faces of the dual lattice, which is also a cubic
lattice.  The faces of the dual lattice correspond to the links in the original
lattice and are also labeled by ${IJ}$.

A membrane operator
is given by
\begin{align}
 Q_a = \sum_{h \in \chi_a} \prod_{IJ \in \text{membrane}} \hat A_{IJ}(h).
\end{align}
where the operator $\hat A_{IJ}(h)$ is defined as
\begin{align}
 \hat A_{IJ}(h) |g_{IJ}\> = |hg_{IJ}\>,
\end{align}
and $\chi_a$ is the $a^\text{th}$ conjugacy class of $G$.  Also $I$'s are on
one side of the membrane and $J$'s are on the other side of the membrane, 


We note that
\begin{align}
\label{QQMQ}
 Q_aQ_b = \sum_{h \in \chi_a} \sum_{\t h \in \chi_b}
\prod_{IJ \in \text{membrane}} \hat A_{IJ}(h\t h)
=\sum_c M^{ab}_c Q_c ,
\end{align}
The above expression allows us to see that
$M^{ab}_c$ are non-negative integers.
Clearly $Q_a Q_b = Q_bQ_a$ and $(Q_a Q_b) Q_c= Q_a (Q_b Q_c)$, which imply that
\begin{align}
 M^{ab}_c &= M^{ba}_c, & \sum_d M^{ab}_d M^{dc}_{e} &= \sum_d M^{ad}_e M^{bc}_{d}
\end{align}
Let $(M_a)_{cb} = M^{ab}_c$, and we can rewrite the second equation in the
above as
\begin{align}
 M_c M_a = M_a M_c .
\end{align}

For example, the permutation group of three elements $S_3 = \{(123),(132),(321),(213),(231),(312)\}$ has
three conjugacy classes: $\chi_1 =\{(123)\}$, $\chi_2 =\{(132),(321),(213)\}$,
and $\chi_3 =\{(231),(312)\}$.  We find that
\begin{align}
Q_1Q_a&=Q_a, & Q_2Q_2&=3Q_1+3Q_3,
\nonumber\\ 
 Q_3Q_3&=2Q_1+Q_3, & Q_2Q_3&=2Q_2.
\end{align}

Let $\v c$ be a common eigenvector of $M_a$ whose components are all
non-negative.  (Such common eigenvector exists since the matrix elements of
$M_a$ are all non-negative.) The eigenvalue of such a eigenvector is $\la_a$
for $M_a$.  We choose the scaling factor of $\v c$ to satisfy
\begin{align}
 \sum_a \la_a c_a =1.
\end{align}
In this case
\begin{align}
 Q^2=Q, \ \ \ \ Q=\sum_a c_a Q_a.
\end{align}

\subsection{A commuting-projector Hamiltonian}

Let $Q_{I,a}$ be the smallest membrane operator that creates a small membrane
corresponding to the surface of a cube in the dual lattice.  Such a membrane
wraps a site $I$ in the original cubic lattice.  We note that $Q_{I,a}$ is a
sum of gauge transformation operators $g_{IJ} \to hG_{IJ}$.  Since the string
operators are gauge invariant, we have
\begin{align}
 [B_i,Q_{I,a}]=0.
\end{align}
Therefore, we can construct the following commuting projector Hamiltonian
\cite{K032,MR1315}
\begin{align}
 H=\sum_I(1-Q_I)+\sum_{\<IJKL\>} (1-B_{\<IJKL\>}),
\end{align}
where
\begin{align}
 Q_I = \sum_a c_a Q_{I,a},\ \ \ \
B_{\<IJKL\>} = \sum_i \frac{d_i}{D} B_{\<IJKL\>,i}
\end{align}
and $\<IJKL\>$ labels the loops around the squares of the original cubic
lattice.

The ground state of the above  exactly soluble Hamiltonian has a nontrivial
topological order.  The low energy effective theory is the $G$-gauge theory.

\subsection{The point-like and string-like excitations}

What are the excitations for the above Hamiltonian?  There are local point-like
excitations created by local operators.  There are also topological point-like
excitations that cannot be created by local operators.  Two topological
point-like excitations are said to be equivalent if they differ by local
point-like excitations.  The equivalent topological point-like excitations are
said to have the same type.  

The different types of topological point-like
excitations are created at the ends of the open string operators that we
discussed before.  Thus we see that types of topological point-like excitations
one-to-one correspond to the irreducible representations of $G$. In other
words, topological point-like excitations are described by $\Rep(G)$ in a
$G$-gauge theory.

Similarly, there are also topological string-like excitations.  They are
created at the boundary of the open membrane operators.  However, the types of
membrane operators are not one-to-one correspond to the types of string-like
excitations.  There are \emph{pure} string-like excitations which one-to-one
correspond to the conjugacy classes of $G$.  There are also \emph{mixed}
string-like excitations which are bound state of pure string-like excitations
and point-like excitations \cite{KW1458,WL1437,EN170202148}.  In general, the types
(pure and mixed) of string-like excitations in a $G$-gauge theory are labeled
by a pair $\chi,R(G_\chi)$, where $\chi$ is a conjugacy class of $G$,
$R(G_\chi)$ is a representation of $G_\chi$, and $G_\chi$ is a subgroup of $G$
whose elements all commute with a fixed element in $\chi$ (a centralizer
subgroup).

For example, in $S_3$-gauge theory, the $\chi_2$-flux-loop breaks the $S_3$
gauge ``symmetry'' down to $Z_2$ gauge symmetry.  So there are two types of
$\chi_2$-flux-loop excitations, one carries no $Z_2$ charge (which is the pure
one denoted by $s_{20}$) and the other carries $Z_2$ charge 1 (denoted by
$s_{21}$).  Similarly,  the $\chi_3$-flux-loop breaks the $S_3$ gauge
``symmetry'' down to $Z_3$ gauge symmetry.  So there are three types of
$\chi_3$-flux-loop excitations, charring $Z_3$ charge $q=0,1,2$ (denoted by
$s_{3q}$).  Thus, the string excitations in $S_3$-gauge theory are given by $
s_{20}=(\chi_2, R_0(Z_2));\ s_{21}=(\chi_2, R_1(Z_2));\ s_{30}=(\chi_3,
R_0(Z_3));\ s_{31}=(\chi_3, R_1(Z_3));\ s_{32}=(\chi_3, R_2(Z_3)) $.  (Note
that $G_{\chi_2}=Z_2$ and $G_{\chi_3}=Z_3$.) Those string like excitations also
have a shrinking rule: if we shrink a string to a point, it will behave like a
point-like excitation:
\begin{align}
\label{shrink}
s_{20} &  \to p_0\oplus p_2,\ \ \ s_{21} \to p_1\oplus p_2,
\nonumber\\
s_{30} &  \to p_0\oplus p_1,\ \ \ s_{31} \to p_2,\ \ \ s_{32} \to p_2.
\end{align}

In general, the string-excitations whose shrinking rule contain the
trivial excitation $p_0$ are the pure string excitations, which are $s_{20}$ and
$s_{30}$ in the $S_3$-gauge example. We see that the types of pure string
excitations are labeled by $\chi$, the conjugacy classes of $G$.

The $\chi_1$-flux-loop (\ie trivial flux-loop) does not break the $S_3$ gauge.
So there are three types of $\chi_1$-flux-loop excitations, carrying a trivial
representation (denoted by $s_{10}$ or $p_0$), a nontrivial 1D irreducible
representation (denoted by $s_{11}$ or $p_1$), or a 2D irreducible
representation (denoted by $s_{12}$ or $p_2$) of $S_3$.  In fact, those
$\chi_1$-flux-loop excitations (or trivial-string excitations) correspond to
the point-like excitations.  The fusions of all those string-like excitations
are given by Table \ref{strfusion}.

We may regard loop-like excitations $(\chi,q)$ with the same
conjugacy class $\chi$ but different representations $q$ as equivalent and introduce a notion of
pure-type: the loop-like excitation $(\chi,q)$ is said to have an
pure-type $\chi$.  So the fusion of the membrane operators correspond to
the fusion of pure-types, which is closely related to the fusion  of
string-like excitations, after we quotient out the $G_\chi$-representations
$q$, by identifying
\begin{align}
 p_0 &=Q_1,\ \ \ p_1 =Q_1,\ \ \ p_2 =2Q_1, 
\nonumber\\
 s_{2q} &=Q_2,\ \ \ s_{3q} =Q_3, 
\end{align}
in the fusion rule table \ref{strfusion}.
The general identification formula is
\begin{align}
 s_{\chi q} =\text{dim}(q) Q_\chi,
\end{align}
where $\text{dim}(q)$ is the dimension of the $G_\chi$-representations $q$.

\section{A general discussion of string and membrane operators in 3+1D topological orders}

\subsection{String operators in 3+1D topological orders}

For a generic 3+1D topological order, the type-$i$ particle-like excitations
are still described by string operators
\begin{align}
 B_i =
\sum_{a_1a_2a_3\cdots} \hat O^{aa_1}_i(I_1) \hat O^{a_1a_2}_i(I_2)
\hat O^{a_2a_3}_i(I_3)\cdots  .
\end{align}
We say that \emph{two closed string operators are equivalent if they differ by
a local unitary transformation.}  More precisely, two closed string operators
are equivalent iff they can deform into each other while keep all the local
operators having short ranged correlations.  We also normalize the string
operators such that $\<B_i\>$ is independent of string length when the string
is closed.  Such normalized string operators satisfy the following fusion
algebra
\begin{align}
\label{BBNB1} 
B_i B_j= N^{ij}_k B_k .
\end{align}
We can show that $N^{ij}_k$ are non-negative integers, by viewing the string
operators as the world-lines in time direction.

\subsection{Membrane operators in 3+1D topological orders}

Similarly, the type-$a$ string-like excitations are described by membrane
operators
\begin{align}
 Q_a =
\sum_{\{a_I\}} \hat O^{a_1a_2a_3a_4}_a(I_1) \hat O^{a_4a_5a_6a_7}_a(I_2)
\hat O^{a_3a_8a_9a_{10}}_a(I_3)\cdots  ,
\end{align}
which is tensor network operator.  Also \emph{two closed membrane operators are
equivalent if they differ by a local unitary transformation.}  More precisely,
two closed membrane operators are equivalent iff they can deform into each
other while keep all the local operators having short ranged correlations.

The equivalent classes of the  membrane operators can be different, for
membrane operators with different topology.  The equivalent classes of the
spherical closed membrane operators, $Q^{S^2}_a$, correspond to the pure
membrane types.  The pure membrane type corresponds to the type for pure
string-like excitations.  For toric closed membrane operators, $Q^{T^2}_a$, the
number of the equivalent classes will in general be different from the number
of the equivalent classes of spherical closed membrane operators.  This is
because toric closed membrane operators may contain closed string operators
wrapping around the non-contractible loops, which generate different equivalent
classes.  Clearly, if we do not have nontrivial point-like excitations, then
there will be a one-to-one correspondence between the spherical membrane
operators and toric membrane operators.


Since the loop of string operator on $S^2$ is always contractible, the
spherical membrane operators does not contain loops of string operator. Thus the
spherical membrane operators are labeled by the conjugacy classes only.  The
spherical membrane operators $Q^{S^2}_\chi$ also satisfy the following fusion
algebra
\begin{align}
\label{QQMQS2} 
Q^{S^2}_{\chi_1} Q^{S^2}_{\chi_2}=\sum_k M^{{\chi_1}{\chi_2}}_{S^2;{\chi_3}} Q^{S^2}_{\chi_3} .
\end{align}
 In particular
\begin{align}
 M^{\chi_1\chi_2}_{S^2;\chi_3} =M^{\chi_1\chi_2}_{\chi_3}
\end{align}
which can be calculated from the fusion of the conjugacy classes $\chi_1$ and
$\chi_2$ (see \eqn{QQMQ}).

%
%

\section{More general properties of
string-like excitations in 3+1D topological orders}\label{appendix:str}

\subsection{Pure string-like excitations and sectors in dimension reduction}

In Section \ref{dimred}, we have shown that in 3+1D, the number of the sectors
$N_1^\text{sec}$ in the dimension reduction is the number of the classes of
string-like topological excitations that can be distinguished by the braiding
with the point-like excitations.
But
in Section \ref{partstrbraid}, we have shown that all string-like
topological excitations can be distinguished from each other via their braiding
properties with the point-like excitations.  Therefore, \myfrm{The number of the
sectors $N^\text{sec}_1$ is the number of type of pure string-like topological
excitations, if all the point-like excitations are bosons.}

%
%

%

Let GSD$_{\sC^{d+1}}(M^d_\text{space})$ be the ground state degeneracy of a
$d+1$D topological order $\sC^{d+1}$ on a closed $d$-dimensional space manifold
$M^d_\text{space}$. We note that
\begin{align}
	\text{GSD}_{\sC^{d+1}}(S^d_\text{space})=1.
\end{align}

Now, let us consider a more general dimension reduction where we reduce
$d$-dimensional space $M^d_\text{space} = L^{d-n}_\text{space}\times S^n$ to
$(d-n)$-dimensional space $L^{d-n}_\text{space}$ by shrinking the $S^n$.
$N^\text{sec}_{n}$ be the number of sectors of the dimension reduced
topological orders.  We find that
\begin{align}
	N^\text{sec}_{n} = \text{GSD}_{\sC^{d+1}}(S^{d-n}\times S^n).
\end{align}
We see that 
\begin{align}
	 N^\text{sec}_n = N^\text{sec}_{d-n}.
\end{align}

\subsection{Pure string-like excitations are labeled by the conjugacy classes
of $G$}

From last section, we see that
for 3+1D AB topological order
\begin{align}
 N^\text{sec}_1 = N^\text{sec}_2 = \text{number of types of pure strings}.
\end{align}
In the following, we like to show that
\begin{align}
 N^\text{sec}_2 = \text{number of types of point-like excitations}.
\end{align}

We consider the GSD on $S^1\times S^2$.  We note that the path integral on
space-time $S^1\times D^3$ gives us a particular ground state on $S^1\times
S^2$.  To obtain other ground states on $S^1\times S^2$ we insert a type-$i$
string operator $B_i$ along the $S^1$ in $S^1\times D^3$. The string operator
is inserted at a particular point on $D^3$.  The insertion of different string
operators generate linear independent states.  This is because the point-like
excitations represented by the string operators have non-degenerate braiding
with the pure string-like excitations.

The braiding between the inserted point-like excitations and the string-like
excitations is described by creating a small loop of strings on $S^2=\prt D^3$.
Then we enlarge the loop and let the loop wrap around the $S^2$.  Such a
braiding process is equivalent to applying the sphere membrane operator $Q_a$
on $S^2=\prt D^3$.  The eigenvalues of the sphere membrane operators $Q_a$
should distinguish all the state created by inserting the string operators
$B_i$.  Therefore, \myfrm{The number of the sectors $N^\text{sec}_2$ in the
$S^2$-dimension reduction of a 3+1D topological order is the number of type of
point-like topological excitations.}
This allows us to show that
\myfrm{ For 3+1D topological orders,  the
number of types of point-like excitations and is the same as the number of
types of pure string-like excitations.}
Moreover, from the fact that the untwisted sector of dimension reduction is the Drinfeld center
$Z[\Rep(G)]$, we know that
\myfrm{ The \emph{pure}
string-like excitations in a generic  3+1D AB topological orders are labeled by
the conjugacy classes of $G$.} 

Now, the dimension reduction of a generic bosonic $3+1$D topological order
$\sC^4$ can be written as
\begin{align}
\label{C4C3}
\sC^4 = \bigoplus_\chi \sC^3_\chi
\end{align}
where $\chi$ is the conjugacy class of the group $G$ whose representations form
the SFC of $\sC^4$, and $\sum_\chi$ sums over all the conjugacy classes of $G$.

From the dimension reduction \eqn{C4C3}, we can also compute
the ground state degeneracy on 3-torus
\begin{align}
 \text{GSD}_{\sC^4}(T^3)
= \sum_\chi
 \text{GSD}_{\sC^3_\chi}(T^2)
\end{align}
Those degenerate ground states form a representation
of the mapping class group of $T^3$, which is $SL(3,\Z)$.

The dimensional reduction leads to reduction of the representation of
$SL(3,\Z)$ to the representations of $SL(2,\Z)$ that characterize the 2+1D
dimension reduced topological orders $\sC^3_\chi$.  We consider $SL(2,\mathbb
Z)\subset SL(3,\mathbb Z)$ subgroup and the reduction of the $SL(3,\mathbb Z)$
representation $R_{\sC^4}$ to the $SL(2,\mathbb Z)$ representations $R_{\sC^3_\chi}$:
\begin{equation}
\label{R3R2}
 R_{\sC^4} = \bigoplus_\chi R_{\sC^3_\chi}.
\end{equation}
The  $SL(3,\mathbb Z)$ representation $R_{\sC^4}$ describes the 3+1D
topological order $\sC^4$ and the  $SL(2,\mathbb Z)$ representations
$R_{\sC^3_\chi}$ describe the 2+1D topological orders $\sC^3_\chi$.  The
decomposition \eqn{R3R2} gives us the dimensional reduction \eqn{C4C3}.

\subsection{String-like excitations are $G$-flux, even in generic 3+1D
AB topological orders}

From the fact that the untwisted sector of dimension reduction is the Drinfeld center
$Z[\Rep(G)]$, we even know that,
\myfrm{The
\emph{pure} string-like excitations in a generic  3+1D AB topological order 
have the same fusion ring as the gauge theory with
the corresponding gauge group $G$.} 
Alternatively, we can argue the above claim using the results for the
string-only boundary.
This is because the bulk string-like
excitations can be obtained by lifting the boundary string-like excitations.
Since a bulk string-like excitation $s_i$ can braid around a boundary
string-like excitation $s^\text{bdry}_{g}$, their fusion satisfy
\begin{align}
 s_i \otimes s^\text{bdry}_g = s^\text{bdry}_g \otimes s_i. 
\end{align}
This allows us to show that
\begin{align}
\label{split}
 s_i = \bigoplus_{g \in \chi} s^\text{bdry}_g \equiv s_{\chi},
\end{align}
where $\chi$ is a conjugacy class of $G$.  Therefore, even in a generic 3+1D
topological order, we may still view string-like excitations as the $G$-gauge
flux which is described the conjugacy classes of the group $G$.  In particular,
the bulk pure string-like excitations fuse like the conjugacy classes (see
\eqn{QQMQ}).  As a result, the quantum dimension of a pure string-like
excitation is given by the size of the conjugacy class: $d_{s_\chi}=|\chi|$.
This is one of the key result of this paper.

There is a simple physical way to understand the relation between the bulk and
boundary string-like excitations.  Since the boundary is induced by the
condensation of all the point-like excitations, it corresponds to breaks all
the ``gauge symmetry''.  So there is no gauge equivalence of $g \sim hgh^{-1}$,
and there is no degeneracy between the flux-loop that induce $g$ monodromy and
$hgh^{-1}$ monodromy.  So, if we bring a bulk string $s_{\chi}$ near the
boundary, it will split $ s_{\chi} \to \bigoplus_{g \in \chi} s^\text{bdry}_g$
(see \eqn{split}).

\section{More about dimension reduction}

We argued that the string-like excitations are $G$-flux, even in generic
3+1D AB bosonic topological orders.
Now we can say more about the 2+1D
topological orders $\sC^3_\chi$ that appear in the dimension reduction
\eqn{C4C3}.  We first note that the point-like excitations in 3+1D topological
order $\sC^4$ are described by $\Rep(G)$ for a group $G$.  In the dimension
reduction, those 3+1D point-like excitations becomes the 2+1D point-like
excitations with trivial mutual statistics between them; they form symmetric
fusion subcategories $\sE_\chi$ of the 2+1D dimension reduced topological orders $\sC^3_\chi$.  

For the conjugacy class
$\chi=\{1\}$, \ie the untwisted sector, 
$\sE_{\chi=\{1\}}=\Rep(G)$ and $\sC^3_{\chi=\{1\}}=Z[\Rep(G)]$
is a minimal modular extension of
$\Rep(G)$.  But what about the other conjugacy classes?

Because $\sC^3_\chi$ is induced by threading a $G$-flux described by
$\chi$ through $S^1$,  such a $G$-flux will break the ``gauge symmetry'' from
$G$ to $G_\chi$,  where $G_\chi$ is a subgroup of $G$ that commutes with a fixed
element in conjugacy class $\chi$.  Therefore, the SFC $\sE_\chi$ in
$\sC^3_\chi$ is given by $\Rep(G_\chi)$.  The 3+1D point-like excitations
described by $\Rep(G)$ will split into 2+1D point-like excitations described by
$\Rep(G_{\chi})$ in each sector  (see Table \ref{Dred}).

Similarly for the other sectors we have \myfrm{The
dimension reduced 2+1D topological orders $\sC^3_\chi$ are minimal modular
extensions of $\Rep(G_\chi)$.}  The \emph{minimal modular extension} means that
the anyons in $\sC^3_\chi$ that are not in  $\Rep(G_\chi)$ all have nontrivial
mutual statistics with the bosons in $\Rep(G_\chi)$.  This condition comes from
the result in Section \ref{partstrbraid}. Note that unlike the untwisted
sector, $\sC^3_\chi$ is in general not the Drinfeld center
of $\Rep(G_\chi)$.



\section{The branching structure of space-time lattice}
\label{branch}

\begin{figure}[t]
\begin{center}
\includegraphics[scale=0.5]{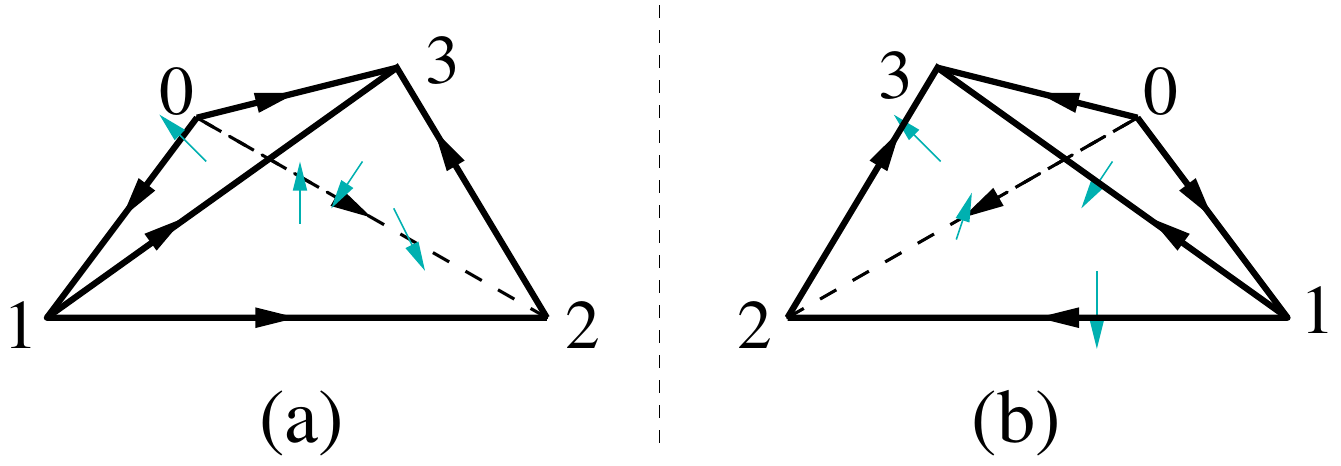} \end{center}
\caption{ (Color online) Two branched simplices with opposite orientations.
(a) A branched simplex with positive orientation and (b) a branched simplex
with negative orientation.  }
\label{mir}
\end{figure}

In order to define a generic lattice theory on the space-time complex
$M^d_\text{latt}$ using local tensors, it is important to give the vertices of
each simplex a local order.  A nice local scheme to order  the vertices is
given by a branching structure.\cite{C0527,CGL1314,CGL1204} A branching
structure is a choice of orientation of each link in the $d$-dimensional
complex so that there is no oriented loop on any triangle (see Fig. \ref{mir}).

The branching structure induces a \emph{local order} of the vertices on each
simplex.  The first vertex of a simplex is the vertex with no incoming links,
and the second vertex is the vertex with only one incoming link, \etc.  So the
simplex in  Fig. \ref{mir}a has the following vertex ordering: $0,1,2,3$.

The branching structure also gives the simplex (and its sub-simplices) a
canonical orientation.  Fig. \ref{mir} illustrates two $3$-simplices with
opposite canonical orientations compared with the 3-dimension space in which
they are embedded.  The blue arrows indicate the canonical orientations of the
$2$-simplices.  The black arrows indicate the canonical orientations of the
$1$-simplices.

\bibliography{./wencross,./all,./publst,local}

\end{document}